\documentclass[graybox, nosecnum]{svmult}

\usepackage{mathptmx}       
\usepackage{helvet}         
\usepackage{courier}        
\usepackage{type1cm}        
\usepackage{makeidx}         
\usepackage{graphicx}        
\usepackage{amsmath}
\usepackage{multicol}        
\usepackage[bottom]{footmisc}
\usepackage{hyperref}        
\usepackage{soul}            
\usepackage{bm}
\usepackage{subfigure}

\newcommand{\nc}{\newcommand}       
\nc{\vc}[1] {\mbox{\boldmath $#1$}} 
\nc{\del}       {\partial}              
\nc{\bra}       {\langle}               
\nc{\ket}       {\rangle}               
\nc{\bras}[1]   {\langle #1|}           
\nc{\kets}[1]   {|#1\rangle}            
\nc{\mapleft}[1]{           
	\smash{\mathop{\,          %
			\hbox to 1.5cm{\rightarrowfill}\, }\limits_{#1}}}
\nc{\beq}     {\begin{eqnarray}} \nc{\eeq}    {\end{eqnarray}}
\nc{\nn}      {\\\nonumber} \nc{\vs}      {\vspace{-0.275cm}}
\nc{\fra}    {\frac{1}{2}}
\nc{\pmat}[1]{\begin{pmatrix}#1\end{pmatrix}}

\makeindex             

\begin{document}
\title*{Pion Exchange Interaction in Bonn Potential and Relativistic and Non-relativistic Framework in Nuclear Matter}
\author{Jinniu Hu \thanks{corresponding author} and Chencan Wang}
\institute{Jinniu Hu \at School of Physics, Nankai University, Tianjin 300071,\\ Shenzhen Research Institute of Nankai University, Shenzhen 518083, China  \email{hujinniu@nankai.edu.cn}
\and Chencan Wang \at School of Physics, Nankai University, Tianjin 300071,  China \email{709887324@qq.com}}
%
%
\maketitle
\abstract{As the residual interaction of quantum chromodynamics in low-energy region, the nucleon-nucleon (NN) potential can only be exactly described by the model picture now. In the Bonn potential, one of the most well-known NN interaction models, the nucleons interact with each other through exchanging the pion and several heavier mesons, where the pion plays an essential role. It provides a partial contribution of tensor force in the intermediate-range region and the main component in the long-range region in NN potential. However, it is very difficult to be treated in the nuclear many-body system due to its pseudovector or pseudoscalar property. Recently, three high-precision charge-dependent Bonn potentials were proposed with pseudovector coupling types and different pion-nucleon coupling strengths and applied them to study the properties of nuclear matter and neutron stars in the non-relativistic and relativistic frameworks. Furthermore, to properly deal with the strong short-range repulsion and tensor force of the NN potential, some new relativistic {\it ab initio} methods have also been developed in the past decade to discuss the role of pion and relativistic effects in nuclear matter.}

\section{Introduction}
Most finite nuclei in the nuclide chart composed of protons and neutrons are complex quantum many-body systems. They are self-bound together with the nucleon-nucleon (NN) potential, which is considered a residual part of one of the four fundamental forces in the universe, with strong interaction at a low-energy scale. In principle, it should be solved by using quantum chromodynamics (QCD) theory. However, it is a long way to directly generate the NN potential from QCD theory due to its unperturbation of the nucleon. The first successful attempt to describe the NN potential with a serious theoretical framework was made by Yukawa. He assumed that the interaction between two nucleons is generated by a particle with a mass of around $100$ MeV, i.e., a pion through analogy with the photon in electromagnetic force~\cite{Yukawa35}.  Finally, the pion was discovered in the cosmic rays~\cite{lattes47}. 

With the development of accelerators, a lot of NN scattering data about proton-proton and proton-neutron was generated at terrestrial laboratories. The basic characters of the NN potential were extracted from the phase-shift analysis. There are strong short-range repulsion and attraction in the intermediate range between two nucleons. When the distance between two nucleons is larger than $2$ fm, their strong interaction becomes weak and is denominated by the pion. Firstly, the NN force model was generated by the free NN scattering information, such as phase shifts, polarizations, and cross-sections in the 1950s, which was called the realistic NN potential. However, in complex nuclei, it was found that the repulsion of NN interaction is much weaker than the realistic NN potential due to the nuclear many-body medium effects. As a result, many effective NN interactions have recently been constructed by reproducing the properties of infinite nuclear matter and finite nuclei~\cite{bender03,stone07,ring96,vretenar05,meng06,niksic11}.  In this paper, the realistic NN potential is mainly discussed.

In the NN scattering, several important symmetries, i.e., rotation invariance, translation invariance, and space reflection invariance, are kept, which should be embodied in the theoretical framework to describe the NN interaction. Furthermore, these scattering processes are strongly spin-dependent. In the 1960s, the NN forces were associated with terms including the spin, momentum, orbital-angular momentum, and coordinate operators, which accord with the above symmetries, like the  Hamada-Johnston potential~\cite{hamada62} and Reid68 potential~\cite{reid68}.

At the same time, several heavier mesons were found in the accelerators, which were adopted to describe the short- and intermediate-region interactions by the Bonn group with one-boson-exchange (OBE) potential~\cite{erkelenz74} based on the quantum field theory. Furthermore, the other baryon degrees of freedom, such as $\Delta$ isobar and the multi-meson exchange terms, were included in the Bonn potential in the 1980s~\cite{machleidt87,machleidt89}. In the 1990s, the charge independence breaking (CIB) and charge symmetry breaking (CSB) effects were considered in NN interaction for the NN scattering phase shift analysis. Therefore, many high-precision NN forces were generated, such as Reid93, Nijmegen 93, Nijmegen I, Nijmegen II, and AV18 potentials~\cite{stoks94,wiringa95}. In 2000, Machleidt proposed a charge-dependent Bonn (CD-Bonn) potential as a high-precision version of the Bonn potential~\cite{machleidt01}. The $\omega, ~\rho, ~\pi$ mesons and two scalar mesons $\sigma_1$ and $\sigma_2$ are exchanged by two nucleons in CD-Bonn potential. It has been widely applied to study the properties of nuclear systems, from light nuclei to heavy nuclei, and infinite nuclear matter. Furthermore, the chiral effective field theory was applied to describe the NN potential, which has been expanded up to the fifth-order expansion~\cite{weinberg90, weinberg91,
	weinberg92,ordonez94,ordonez96,epelbaum98,epelbaum00,entem03,epelbaum05,entem15,epelbaum15a,epelbaum15b,entem17,reinert17,lu22}.    

Due to the strong repulsion of the NN potential at a short-range distance, which was first proposed by Jastrow from abundant experimental data of NN scattering~\cite{jastrow51}, the nuclear many-body system cannot be described by the NN potential with the perturbation theory. Furthermore, the tensor force mainly generated by the pion is also very difficult to include due to its non-central character. Both of them can raise the strong high momentum correlations between two nucleons in the nuclei.

Therefore, the realistic NN potential should be renormalized at low momentum in the nuclear medium to generate a softer interaction so that finite nuclei and symmetric nuclear matter can form the bound states. The earliest NN renormalization scheme was given by Brueckner et al., where the strong repulsion is reduced by the Bethe-Goldstone equation~\cite{brueckner54,bethe56}. Meanwhile, Jastrow exhibited another way to treat the high momentum correlations with a variational method~\cite{jastrow55}. Besides these two schemes, many relevant nuclear many-body methods with realistic NN potential to calculate the finite nuclei and infinite nuclear matter have been developed, i.e., {\it ab initio} method, such as the Brueckner-Hartree-Fock (BHF) method~\cite{li06,baldo07,baldo16}, lowest order constrained variational method~\cite{modarres93}, quantum Monte Carlo methods~\cite{akmal98,carlson15}, self-consistent Green's function method~\cite{dickhoff04}, coupled-cluster method~\cite{hagen14a,hagen14}, many-body perturbation theory~\cite{carbone13,carbone14,drischler14}, functional renormalization group (FRG) method~\cite{drews15,drews16}, and so on. 

However, most of the nuclear {\it ab initio} methods were constructed in the non-relativistic framework, which cannot completely reproduce the empirical nuclear saturation properties, $E/A=-16\pm1$ MeV at $n_0=0.16\pm0.01$ fm$^{-3}$ and the ground state properties of finite nuclei only with the present two-body realistic nuclear force. The additional three-body nuclear potential must be included~\cite{li06,hu17,sammarruca18,logoteta19}.

In the 1980s, Anastasio et al. proposed the relativistic version of the BHF model, i.e.,  the relativistic Brueckner-Hartree-Fock (RBHF) method~\cite{ansatasio83}. It was developed later by Horowitz {et al.}~\cite{horowitz87} and Brockmann et al.~\cite{brockmann90}. A repulsive contribution is provided due to the relativistic effect in the RBHF model, which can reproduce the empirical nuclear saturation properties with the Bonn potential. It was explained that the nucleon-antinucleon excitation in the relativistic effect has a similar contribution to that of the three-nucleon force~\cite{lizh08}. Furthermore,  the superfluity of nuclear matter, properties of the neutron star, and nuclear density functional theories were studied within the RBHF model~\cite{alonso03,krastev06,sammarruca10,dalen10}. Recently, a fully self-consistent calculation of the RBHF model in finite nuclei system and extended this framework on the neutron drops were realized~\cite{shen16,shen17,shen19}. 

However, there are still many unsolved problems in the relativistic {\it ab initio} method. The nucleon should be considered as a Dirac particle in the relativistic framework. Therefore, nucleon-nucleon potentials must be constructed in the quantum field theory. There are only Bonn and CD-Bonn potentials until 2018, which can be used in the RBHF model. However, the Bonn potential was generated at the beginning of 1990. It does not consider the CIB and CSB effects and is not a high-precision NN potential. Furthermore, the coupling between pion and nucleon in the CD-Bonn potential is the pseudoscalar type, which produces an additional attractive contribution~\cite{fuchs98}. Hence, it is urgent to develop a high-precision NN potential that can be adopted by the RBHF model.

On the other hand, the normalization schemes from the Brueckner theory cannot clearly distinguish the roles of short-range correlation and tensor correlation from the pion. It is necessary to develop the new relativistic   {\it ab initio} beyond the mean-field approximation, which can treat the strong repulsion at short-range distance and attraction at intermediate range from pion with realistic NN potential.

In this paper, our recent progress in the high-precision NN potential and relativistic {\it ab initio} method will be shown. In Sect. II, the revised CD-Bonn potentials with pseudovector coupling and different tensor components are given. In Sect. III, the properties of nuclear matter and neutron stars with the revised CD-Bonn potential will be discussed within the RBHF model. In Sect. IV, several new relativistic {\it ab initio} methods are shown beyond the relativistic Hartree-Fock model. In the last section, the summary and perspective will be given.

\section{Charged-dependent Bonn potential with pseudovector coupling}
The behavior of NN interaction in the long-range region is mainly denominated by the pion, which also represents the chiral symmetry of QCD theory as a Goldston boson, but the coupling types between nucleon and pion have two schemes, i.e., the pseudoscalar (PS) and pseudovector (PV) coupling. The on-shell amplitudes of PS and PV coupling between pion and nucleon are completely the same. 

However,  the PV coupling was preferred in the calculations of pion-nucleon scatterings, such as $\pi N$ to the $\pi\pi\rightleftharpoons N\bar N$ and so on~\cite{lacombe75,jackson75}. In the chiral perturbation theory, the PV type was also taken in low-energy region to investigate the pion electroproduction and photoproduction~\cite{drechsel92,drechsel99}. On the other hand, an additional strong attraction was obtained for its strong coupling to negative energy states, when the NN potentials including the PS coupling were applied to the relativistic framework. Instead, the PV coupling can suppress this contribution due to the vanishment of the matrix element between the antinucleon and nucleon in the on-shell scattering~\cite{fuchs98}. 

The high-precision CD-Bonn potential proposed by Machleidt~\cite{machleidt01} with the analogous framework of Bonn potential should be adopted in the RBHF model. However, its coupling scheme between pion and nucleon is the PS type. Therefore, in a relativistic framework, it will provide a very attractive contribution and the empirical saturation properties thus can not be reproduced. Therefore, the PV coupling is adopted instead of the PS one in the original CD-Bonn potential. 

In the CD-Bonn potential, five mesons are considered to be exchanged in the NN interaction. Two scalar mesons are $\sigma_1$ and $\sigma_2$. Two vector mesons are $\omega$ and $\rho$. The pion is recognized as the PV meson. The Lagrangian between them and nucleons can be written as, 
\beq
	\mathcal{L}_{\pi^0 NN}&=&  -{f_{\pi NN}\over m_{\pi^\pm}}
	\bar{\psi}\gamma^5\gamma^\mu  \psi \cdot\partial_\mu\phi^{(\pi^0)},\nn
	\mathcal{L}_{\pi^\pm NN}&=&  -{\sqrt{2}f_{\pi NN}\over m_{\pi^\pm}}
	\bar{\psi}\gamma^5\gamma^\mu \tau_\pm \psi \cdot\partial_\mu\phi^{(\pi^\pm)},\nn
	\mathcal{L}_{\sigma NN}&=&  -g_{\sigma NN}\bar{\psi} \psi \phi^{(\sigma)},\nn
	\mathcal{L}_{\omega NN}&=&  -g_{\omega NN}\bar{\psi} \gamma^\mu \psi \phi^{(\omega)}_\mu,
	\nn
	\mathcal{L}_{\rho  NN}&=&  -g_{\rho NN}
	\bar{\psi} \gamma^\mu  \vec{\tau} \psi  {\vec{\phi}}_\mu^{(\rho)}
	-{f_{\rho NN}\over 2M_p}\bar{\psi} \sigma^{\mu\nu}
	\vec{\tau}\psi \partial_\mu \vec{\phi}^{(\rho)}_\nu,
\eeq
where $\psi$ and $\phi$ denote the nucleon and meson fields, respectively. $\vec{\tau}$ is the isospin operator. $g$ and $f$ are the coupling constants. $m_\pi$ and $M_p$ are the masses of pion and nucleon, respectively.

The NN interaction can be obtained by the Feynman scattering amplitude between meson and nucleon,
\begin{equation}\label{ch2-obep}
	\bar{V}_\alpha(\bm{q}',\bm{q}) = -
	\bar{u}_1(\bm{q}')\Gamma_\alpha^{(1)} u_1(\bm{q})
	{iP_\alpha  \over \bm{k}^2 + m_\alpha^2}
	\bar{u}_2(-\bm{q}')\Gamma_\alpha^{(2)} u_2(-\bm{q}),
\end{equation}
where $u_i$ is the spinor at free space,
\begin{equation}\label{ch2-spinor}
	u_i (\bm{q}) = \sqrt{E+M\over 2M}
	\pmat{ 1  \\ {\bm{\sigma}_i\cdot\bm{q}\over 2M}},\quad
	\bar{u}_i (\bm{q}') = \sqrt{E'+M\over 2M}
	\left( 1,~ -{\bm{\sigma}_i\cdot\bm{q}'\over 2M}\right).
\end{equation}
$E=\sqrt{\bm{q}^2+M^2}$ and $E'=\sqrt{{\bm{q}'}^2+M^2}$ are the input and output energies of on-shell nucleon. $\Gamma^{(i)}_\alpha$ is the coupling vertex between meson and nucleon. $\bm{q}'$ and $\bm{q}$ are the corresponding momenta. $iP_\alpha/(\bm{k}^2+m_\alpha^2)$ is the meson propagator without retarded effect.  $\bm{k}=\bm{q}'-\bm{q}$ is the transferring momentum. Therefore, the NN interaction from various mesons can be summarized as
\begin{equation}\label{ch2-vobep}
	V(\bm{q}',\bm{q}) =  \sqrt{M\over E} \sqrt{M\over E'}
	\sum_{\alpha=\pi^0,\pi^\pm,\omega,\rho,\sigma}
	\mathcal{F}_\alpha(\bm{k}^2) \bar{V}_\alpha(\bm{q}',\bm{q}),
\end{equation}
where, $\sqrt{M\over E} \sqrt{M\over E'}$ comes from the minimal relativity. Furthermore, a form factor to denote the size of nucleon must be included in the vertex between meson and nucleon, \begin{equation}\label{ch2-formfact}
	\mathcal{F}_\alpha(\bm{k}^2) = \left({\Lambda_\alpha^2 -m_\alpha^2\over
		\Lambda_\alpha^2+\bm{k}^2}\right)^{n_\alpha}
\end{equation} 
to regularize the high momentum behaviors of NN interaction. $\Lambda_\alpha$ is called the momentum cut-off and in Bonn potential, $n_\alpha=2$ is the dipole case. In the NN scattering and infinite nuclear matter, the partial wave presentation is convenient
$\bra{\ell'  sj|V(q',q)|\ell  sj}\ket$, which is transformed from the  helicity presentation by 
\begin{equation}
	\bra{\lambda_1'\lambda_2'|V(\bm{q}',\bm{q})|\lambda_1\lambda_2}\ket
	\rightarrow
	\bra{\lambda_1'\lambda_2'|V_j(q',q)|\lambda_1\lambda_2}\ket
	\rightarrow
	\bra{\ell'  sj|V(q',q)|\ell  sj}\ket
\end{equation} 

\begin{table}
	\centering
	\caption{Three kind values of $g_\pi^2/4\pi$ and cut-offs $\Lambda_\pi$, which are named as A, B, and C.}
\begin{tabular}{ccccc}
	\hline\hline
	& $m_{\pi^0}$  [MeV] & $m_{\pi^\pm}$ [MeV] & $g_\pi^2/4\pi$  &
	$\Lambda_\pi$ [GeV]  \\
	\hline
	A  & 134.9766   & 139.5702  & 13.9 & 1.12  \\
	B  & 134.9766   & 139.5702  & 13.7 & 1.50  \\
	C  & 134.9766   & 139.5702  & 13.6 & 1.72\\
	\hline   \hline
\end{tabular}
\label{tab1}
\end{table}

Three pion-nucleon coupling strengths are considered to discuss the role of pion in NN interaction as shown in Table~\ref{tab1}~\cite{wang19}. Due to the largest cut-off in the C case, its have the strongest pion contribution in NN potential. For the $\omega$ and $\rho$ mesons, their coupling constants with nucleon and curoffs are taken as the conventional magnitudes from the hadron physics given in Table~\ref{tab21}.
\begin{table}[htb]
	\centering
	\caption{The parameters of vector mesons, including their mass, $m$, coupling strengths, $g^2/4\pi$, and cut-offs $\Lambda$.}
	\begin{tabular}{cccccc}
		\hline\hline
		mesons &~$m$ [GeV]~&   ~$g^2/4\pi$  ~& ~$f/g$ ~& ~$\Lambda$ [GeV]\\
		\hline
		$\rho^0,~\rho^\pm$ &  0.770  &  0.84  &  6.1 & 1.31 \\
		$\omega$&0.783 &  20  &   0    & 1.50 \\
		\hline   \hline
	\end{tabular}
\label{tab21}
\end{table}

\begin{figure}[htb]
	\centering
	\includegraphics[width=0.9\linewidth]{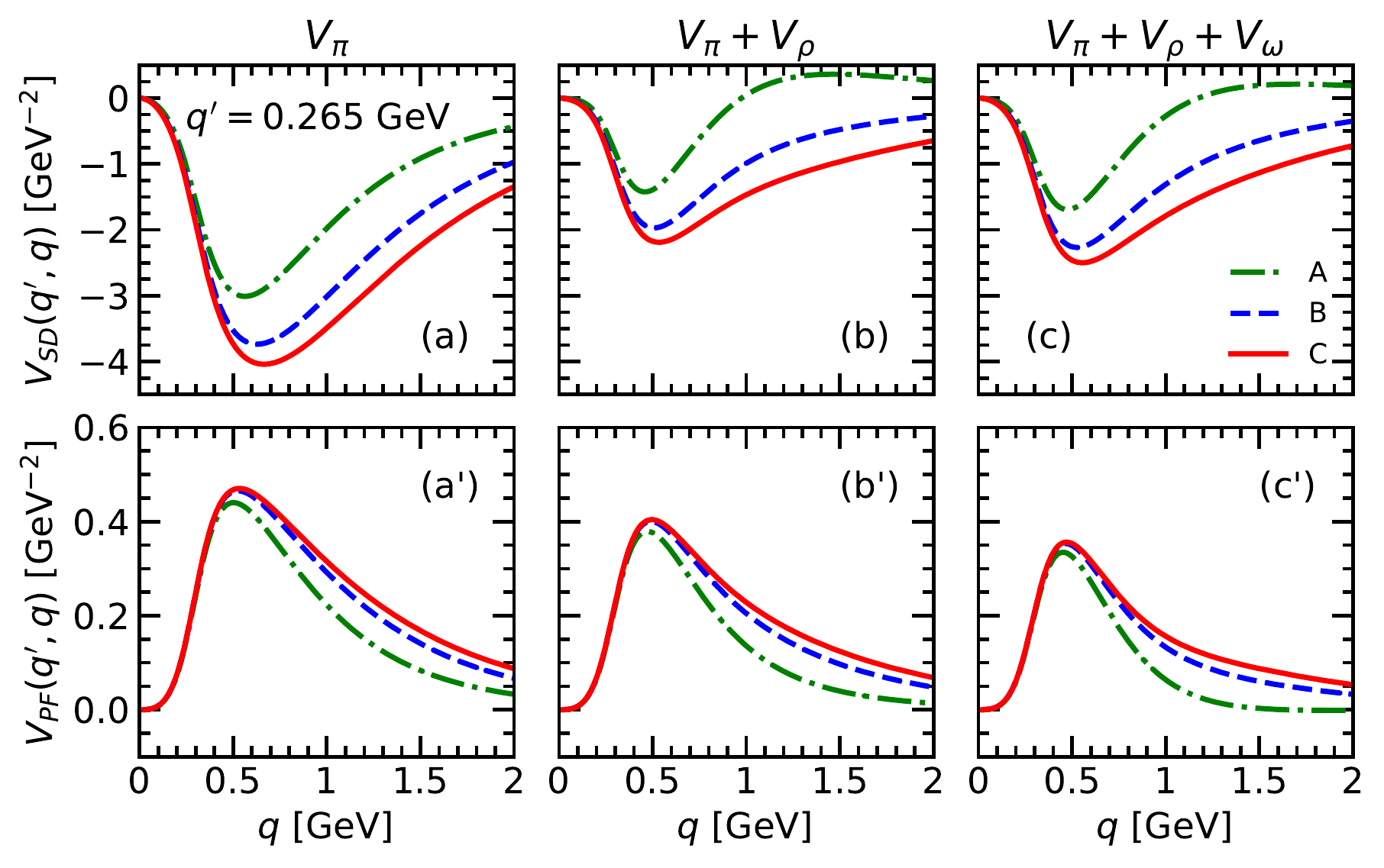}
	\caption{The tensor contributions from the pion and vector mesons in $^3S_1 - {}^3D_1$ (upper panels) and $^3P_2-{}^3F_2$ channels (lower panels). In panel (a), only pion is included. In panel (b), pion and $\rho$ meson are considered. In panel (c), all mesons are taken into account. Here, the output momentum is fixed as $q'=0.265$~GeV.}
	\label{fig1}
\end{figure}

The pion, $\omega$, and $\rho$ mesons all can generate the tensor force, which is denoted as the matrix elements between the $\ell$ and $\ell+2$ states, such as $S-D$ or $P-F$ channels.  In Fig.~\ref{fig1}, the half-on-shell matrix elements at $^3S_1 - {}^3D_1$ and$^3P_2-{}^3F_2$ channels are shown with a fixed output momentum $q'=0.265$~GeV. Pion provides the most attractive contributions at $S-D$ channels. When the $\rho$ meson is included, the magnitudes are largely reduced. Meanwhile, the $\omega$ meson can also generate some attractive components. In the $P-F$ channel, the matrix elements of pion are positive, while those from the $\rho$ and $\omega$ mesons are negative, which are strongly dependent on the isospin characters of NN potential.

The parameters about the scalar mesons, $g_{\sigma_1}^2/4\pi,~m_{\sigma_1},~g_{\sigma_2}^2/4\pi,~m_{\sigma_2}$ are determined by the partial wave phase shifts at each channel extracted from the NN scattering and the binding energy of deuteron. In the center of mass system (c.m.), the scattering process is described by the Lippmann-Schwinger (LS) equation,
\begin{equation}\label{ch2-lseq}
	T(\bm{q}',\bm{q}) = V(\bm{q}',\bm{q})
	+ \int {{d}^3 k\over (2\pi)^3}V(\bm{q}',\bm{k})
	{M\over \bm{q}^2 - \bm{k}^2 + i\epsilon}T(\bm{k},\bm{q}).
\end{equation}
In the partial wave basis $|\ell s j\rangle$ with the conservations of total spin and angular momentum, this equation can be expanded as 
\begin{equation}\label{ch2-lseq-pw}
	T_{\ell'\ell s j}(q',q) = V_{\ell'\ell s j}(q',q) +
	\sum_{\ell''}\int k^2{{d}k}~V_{\ell'\ell'' s j}(q',k)
	{M\over q^2 -k^2 + i\epsilon}T_{\ell''\ell s j}(k,q),
\end{equation}
where, $V_{\ell'\ell  sj}=\bra{\ell' js|V|\ell js}\ket$. The phase shift, $\delta_j$ can be related to the $T$ matrix through the $S$ matrix. The parameters about two scalar mesons are obtained by minimizing the function with the least square method,
\begin{equation}\label{ch2-lossfunc}
	\chi^2 = {1\over N}\sum_{i=1}^N \left(\delta  -\delta_{Nijm} \right)_i^2.
\end{equation}
The input energies of nucleon at laboratory frame, $E_{lab}$ are less than $300$ MeV, here. $\delta_{Nijm}$ is the Nijmegen partial wave analysis (PWA). First, the proton-proton (pp) potential is generated by this fitting process. Then, for the proton-neutron (np) and neutron-neutron (nn) potentials, the CSB and CIB effects are considered. Finally, three high-precision charge-dependent Bonn potentials with PV coupling are obtained, which can nicely reproduce the data from Nijmegen PWA, i.e., pvCDBonn A, B, and C potentials.

\begin{figure}[htb]
	\centering
	\includegraphics[width=1\linewidth]{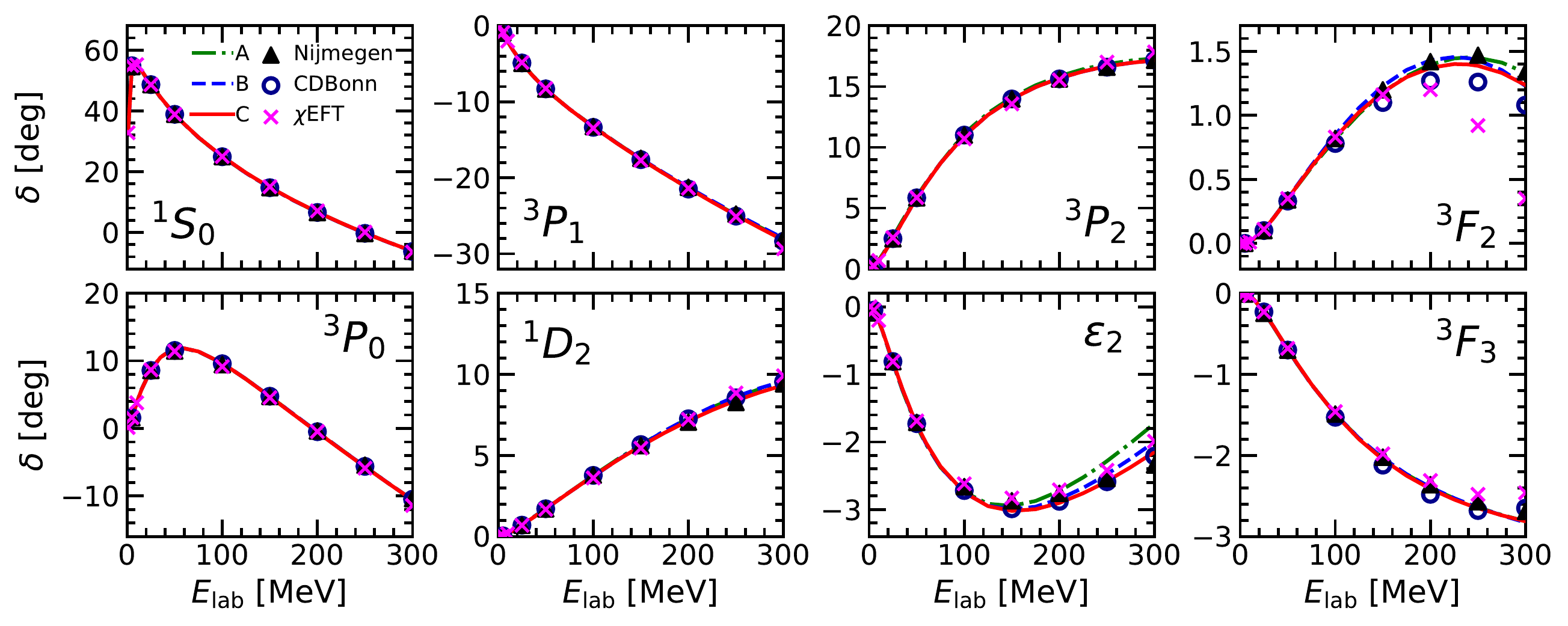}\\
	\caption{The phase shifts of pp from pv CDBonn A, B, and C potentials and compared to those from the Nijmegen PWA, origin CD-Bonn potential and latest chiral potential.}
	\label{fig2}
\end{figure}
In Fig.~\ref{fig2}, the phase shifts of pp scattering from pvCDBonn A, B, and C potentials are shown and compared to those from the Nijmegen PWA, original CD-Bonn, and latest chiral NN potential at different channels. It can be found that all of them can nicely reproduce the data from the Nijmegen PWA and are consistent with the results from the other high-precision NN potential. The only difference appears in the mixing parameter $\epsilon_2$ at $^3P_2-^3F_2$ channel, which is related to the strengths of tensor components in these three potentials. There is a similar situation for the np scattering at the $^3S_1-^3D_1$ channel. 

\begin{figure}[htb]
	\centering
	\includegraphics[width=1\linewidth]{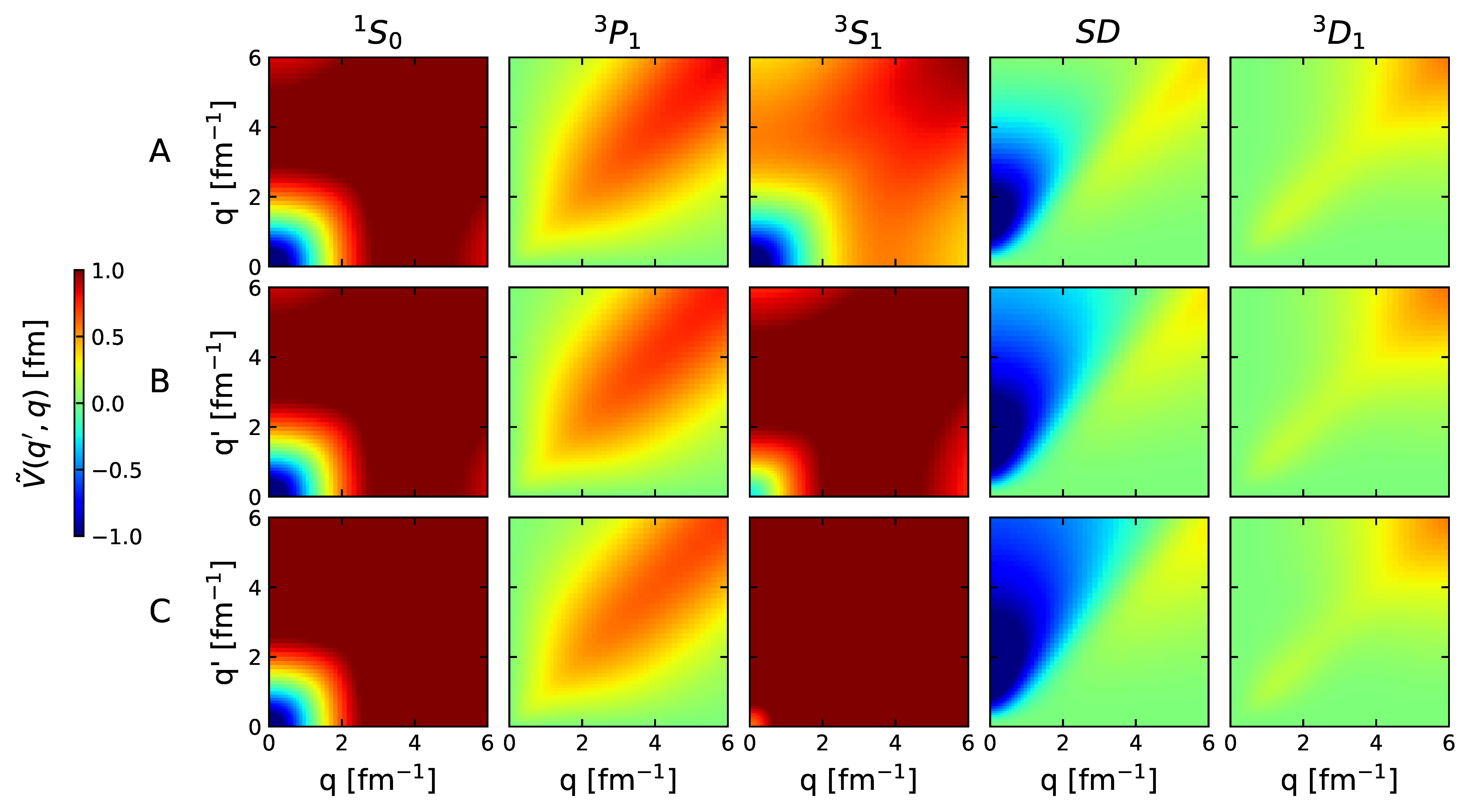}\\
	\caption{The matrix elements of np components from pvCDBonn potentials. They are rescaled with $\tilde{V}=M\pi V/2$ whose unit is fm.}
	\label{fig3}
\end{figure}
In Fig.~\ref{fig3}, the matrix elements of np components from pvCDBonn A, B, and C potentials are shown in the two-dimensional contour as functions of momenta. It is clear that most differences among them at $^3S_1$ and $^3S_1-^3D_1$ channels, which are denominated by the pion. At the low momentum region, there is attractive interaction at $^3S_1$ channel for pvCDBonn A potential, while they become the repulsive one at pvCDBonn C potential, which has the strongest pion contribution. Meanwhile, the situation is the opposite at the $^3S_1-^3D_1$ channel. The pvCDBonn C has the strongest attractive contribution in the low-moment region. Therefore, the role of the pion in the NN potential is mainly embodied in the coupled channels.

When the relative momentum $q$ is very small, the phase shift at the $S$ wave can be denoted by the scattering length, $a$ and effective range, $r$,
\begin{equation}
	q\cot \delta_S \approx -{1\over a} + {1\over 2} rq^2 + \mathcal{O}(q^4),
\end{equation} 
which can be measured from the experiments. Their values from the pvCDBonn potential are shown in Table~\ref{tab2}, which are consistent with results from the original CD-Bonn potential and the empirical data from experiments.

\begin{table}[htb]
	\centering
	\caption{The scattering length ($a$) and effective range ($r$) for pvCDBonn A, B, and C potential and the empirical data from experiments. Their units are fm. $a_{t}({np})$  and $r_{t}({np})$  correspond the scattering length and effective range from spin triplet channel at $^3S_1$ state.}
\begin{tabular}{cccccc}
	\hline\hline
	&  A & B & C & CD-Bonn~\cite{machleidt01} & Emp. \\
	\hline
	$a({nn})$  & $-$18.936& $-$18.931 & $-$18.917 & $-$18.968 & $-$18.95(40)
	\cite{howell98,gonzalez99}  \\
	$r({nn})$  & 2.778  &  2.770  &  2.765  & 2.819  &2.75(11)
	\cite{miller90} ~~~~~~\\
	$a({np})$  & $-$23.711& $-$23.757 & $-$23.752 &$-$23.738 & $-$23.749(20)
	\cite{houk71,machleidt01}\\
	$r({np})$  & 2.649  & 2.649   & 2.642   & 2.671  &2.77(5)
	\cite{houk71,machleidt01}
	\\
	$a_t({np})$& 5.429  & 5.418   & 5.420   &5.420 & 5.419(7)
	\cite{houk71,machleidt01}\\
	$r_t({np})$& 1.773  & 1.757   & 1.761   &1.751 & 1.753(8)
	\cite{houk71,machleidt01}\\
	\hline
	\hline
\end{tabular}
\label{tab2}
\end{table}

The lightest nuclear bound state, deuteron can be solved with a on-shell LS equation,
\begin{equation}\label{ch2-deutwaveq}
	\Psi(q) = -{M\over \gamma_{d}^2 + q^2}\int_0^{+\infty} q'^2 {d}q'
	~V(q,q')\Psi(q'),  
\end{equation}
where $ \gamma_{d}^2/M=B_{d}$ is the bound energy of deuteron and $M$ the mean mass of proton and neutron. The wavefunction $\Psi$ should have two components at $S$ and $D$ waves. The properties of the deuteron, the asymptotic $S$-state amplitude ($A_S$),  the ratio of the $D/S$ states ($\eta$),  the root-mean-square radius of the deuteron ($r_d$), the quadrupole moment ($Q_d$), and the $D$-state probability ($P_D$) from pvCDBonn potentials, original CD-Bonn potential and experiment data are given in Table~\ref{tab4}. They are consistent with the experiment data. The $D$-state probability represents the strength of the tensor component in an NN potential. Therefore, there is the strongest tensor component in the pvCDBonn C potential.

\begin{table}[htb]
	\centering
	\caption{The properties of deuteron from pvCDBonn potentials, original CD-Bonn potential and experiment data.}
\begin{tabular}{cccccc}
	\hline\hline
	&  A & B & C & CD-Bonn~\cite{machleidt01}& Exp. \\
	\hline
	$A_S$ [fm$^{-1}$]   & 0.8818 &  0.8828  & 0.8856  & 0.8846 & 0.8846(9)
	\cite{kermode83} \\
	$\eta$              & 0.0237& 0.0246  & 0.0250 & 0.0256& 0.0256(4)
	\cite{rodning90}\\
	$r_{d}$   [fm]   & 1.970 & 1.967   & 1.969  & 1.966 & 1.97507(78)
	\cite{martorell95} \\
	$Q_{d}$ [fm$^2$] & 0.258 & 0.269   & 0.275  & 0.270 & 0.2859(3)
	\cite{bishop79}   \\
	$P_D$ [\%]          & 4.279 & 5.493   & 6.152  & 4.850 &   \\
	\hline
	\hline
\end{tabular}
\label{tab4}
\end{table}

\section{The nuclear matter in BHF and RBHF models}
The infinite nuclear matter is an idea object consisting of protons and neutrons only with the NN interaction. To avoid the divergence of energy, the coulomb interaction between protons is neglected. For symmetric nuclear matter with the same number of protons and neutrons, its saturation properties, such as saturation density and binding energy, can be extracted from the center region of heavy nuclei. 

The properties of nuclear matter, which is not only dependent on the density but also on the fractions of proton and neutron, especially in the high-density region, are very important for the compact star and supernova explosion. There are several important properties to describe nuclear matter, i.e. binding energy per nucleon ($E/A$), nuclear saturation density ($n_0$) the symmetry energy ($E_{sym}$), the slope of symmetry energy at $n_0$ ($L$), the incompressibility ($K$), and the effective nucleon matter ($M^*_N$).

The nuclear matter can be investigated both by the density functional theories \cite{bender03,stone07} and the {\it ab initio} methods~\cite{li06,dickhoff04,hagen14,carbone13,drischler14,hu17}. However, there are large uncertainties in the high-density region for binding energy per nucleon from density functional theories~\cite{li19}. As a result, it is more reasonable to treat the nuclear matter with {\it ab initio} methods. In this section, the properties of nuclear matter will be studied in BHF and RBHF models.

In the non-relativistic framework, the nucleon moving in the nuclear matter is given by a plane wave function, $\psi\sim \exp(-i\bm{p\cdot x})$. In the relativistic framework, the nucleon is described by the Dirac equation,
\begin{equation}\label{ch3-diraceq1}
	(p\!\!\!\!/ -M -\Sigma)u(\bm{p},\lambda)=0,
\end{equation} 
 where, 
 \begin{equation}\label{ch3-se}
 	\Sigma=\Sigma_{s}-\gamma^0\Sigma_0 + \bm{\gamma}\cdot\bm{p}\Sigma_{v}
 \end{equation}
is the self-energy of the nucleon in the nuclear medium. It has three components, scalar one $\Sigma_s$, time component, $\Sigma_0$, and  $\Sigma_v$, which are density-, momentum-, and isospin-dependent. The effective mass and energy are defined as,
\begin{equation}\label{ch3-effq}
	M^* = {M+\Sigma_{s}\over 1+ \Sigma_{v}},\quad
	E^* = {E+\Sigma_{0}\over 1+ \Sigma_{v}}.
\end{equation}
Therefore, the Dirac equation about the nucleon in nuclear medium has a solution,
\begin{equation}
	E^*(\bm{p}) = \sqrt{\bm{p}^2 + M^{*2}},\quad \
	u(\bm{p},\lambda) =\sqrt{E^* + M^*\over 2M^*} \pmat{1 \\
		{\bm{\sigma}\cdot \bm{p}\over M^* + E^*}} |{\lambda}\ket.
\end{equation}

The normalization of the NN potential in Brueckner theory is based on the Bethe-Brueckner-Goldstone (BBG) equation, which is reduced from the Bethe-Salpeter equation in the nuclear medium from four-dimension to three-dimension with Blankenbecler-Sugar choice,
\begin{equation}\label{ch3-Gbar}
	\begin{aligned}
		{G}_{\tau\tau'}(\bm{q}',\bm{q}|x) &=& {V}_{\tau\tau'}(\bm{q}',\bm{q}) +
		\int{{d}^3k\over(2\pi)^3}{V}_{\tau\tau'}(\bm{q}',\bm{k})
		{M^*_\tau M^*_{\tau'} \over E^*_\tau(\bm{k}) E^*_{\tau'}(\bm{k})}\nn
		&&\times{2W_k \over W_0 + W_k} {Q_{\tau\tau'}(\bm{k},x) \over W_0 -W_k +i\epsilon}
		{G}_{\tau\tau'}(\bm{k},\bm{q}|x).
	\end{aligned}
\end{equation}
Here, ${G}_{\tau\tau'}$ is the effective NN potential in nuclear medium $x=\{W_0,~\bm{P},~\bm{u},~k_F^\tau,~k_F^{\tau'}\}$. $\tau$ denotes the isospin degree of freedom of nucleon. $\bm{P}=\bm{p}_\tau+\bm{p}_{\tau'}$ is the total momentum and $\bm{u}=\bm{P}/( E_\tau^* + E_{\tau'}^*)$  is the relative velocity of the c.m. frame to the rest frame of nuclear matter. ${V}$ is the realistic NN potential. $Q$ is the Pauli operator, which prevents the states scattering from above Fermi momentum states to the occupied states.

\begin{figure}[hbt]
	\centering
	\includegraphics[width=0.8\linewidth]{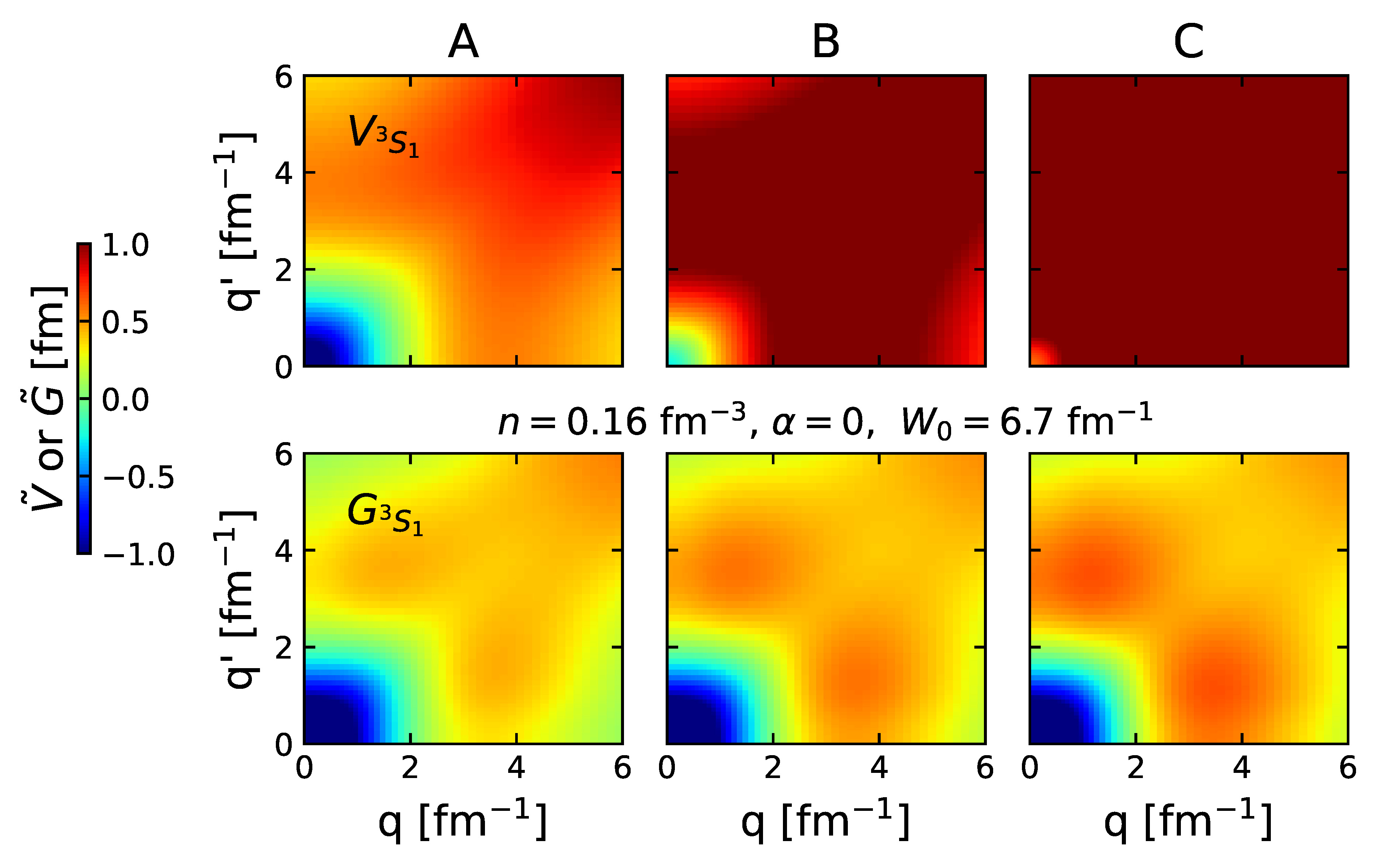}\\
	\caption{ The G matrix elements of pvCDBonn A, B, and C potentials at $^3S_1$ channel with BBG
	equation and compared to the realistic ones, where $n=0.16$~fm$^{-3}$ and the starting energy is $W_0=6.7$~fm$^{-1}$. This unit of this figure is same as Fig.~\ref{fig3}.}
	\label{fig4}
\end{figure}

The matrix elements of G-matrix from pvCDBonn A, B, and C potentials at $^3S_1$ channel are plotted and compared to the realistic ones in Fig.~\ref{fig4}. With the normalization of the BBG equation, the pvCDBonn potentials are much softer. The very strong repulsion of pvCDBonn C potential is completely removed at the low-moment region and transforms into an attractive contribution. In the high-density region, the repulsive components are also weakening.   

With the G-matrix, the binding energy per nucleon in nuclear matter,  $E_{B}/A$ can be obtained based on the Hartree-Fock approximation as
\begin{equation}
	\begin{aligned}
		{E_\mathrm{B}\over A}(n,\alpha) & = E_\mathrm{kin} + E_\mathrm{pot},\nn
		E_\mathrm{kin} & = {1\over n}\sum_{\tau,\lambda}
		\int^{k_F^\tau} {{d}^3p\over (2\pi)^3}
		{M_\tau^*\over E_\tau^*(\bm{p})}
		\bra{{\bar u}(\bm{p},\lambda)|p\!\!\!\!\slash+M_\tau|
			u(\bm{p},\lambda)}\ket-{\overline M}, \nn
		E_\mathrm{pot} &= {1\over 2n}
		\sum_{\tau,\tau'}
		\sum_{\lambda,\lambda'}\int^{p<k_F^\tau}{d^3 p\over (2\pi)^3 }
		\int^{p'<k_F^{\tau'}}
		{d^3 p'\over (2\pi)^3 }{M^*_{\tau}\over E^*_{\tau}(\bm{p})}
		{M^*_{\tau'}\over E^*_{\tau'}(\bm{p}')}
		 \nn
		&\times
		\langle \bar{u}(\mathbf{k},\lambda)\bar{u}(\mathbf{k}',\lambda')
		|{ G}|u(\mathbf{k},\lambda)u(\mathbf{k}',\lambda')\rangle,
	\end{aligned}
\end{equation}
where $E_\mathrm{kin}$ is the energy contribution from kinetic energy and $E_\mathrm{pot}$ from the potential energy. ${\overline M}=[(1-\alpha)M_{p}+(1+\alpha)M_{n}]/2$ is the average nucleon mass. $\alpha$ is the asymmetry factor related to the proton and neutron densities, $\alpha = (n_{n}-n_{p})/(n_n+n_p)$. The Fermi momentum $k_F$ is calculated with the Fermi gas model,
\begin{equation}
	k_F^\tau = \left(3\pi^2 n_\tau \right)^{1\over 3},\quad
	n_\tau = {{k_F^\tau}^3\over 3\pi^2}.
\end{equation}
The effective nucleon mass is a key point to realize the self-consistent BHF calculations, no matter whether in a non-relativistic or relativistic framework. It can be extracted from the self-energy of nucleon or the single-particle potential,
	 \begin{equation}\label{ch3-usp}
		U_{\tau,{sp}}(|\bm{p}|) = \sum_{\tau',\lambda'}
		\int {{d}^3 p'\over (2\pi)^3}	
		{M^*_{\tau'}\over E^*_{\tau'}(\mathbf{p}')}
		\bra{\bar{u}_\tau(\bm{p},\lambda)\bar{u}_{\tau'}(\bm{p}',\lambda')
			|{ G}_{\tau\tau'}|
			u_\tau(\bm{p},\lambda)u_{\tau'}(\bm{p}',\lambda')}\ket,
	\end{equation}
which are named the projection method~\cite{fuchs98}, or fitting method~\cite{brockmann90}, respectively.	In the BHF model, the effective mass just appears at the single particle energy, while in the RBHF model, it will arise in the spinor structure of the Bonn potential and BBG equation.

\begin{figure}[htb]
	\centering
	\includegraphics[width=0.85\linewidth]{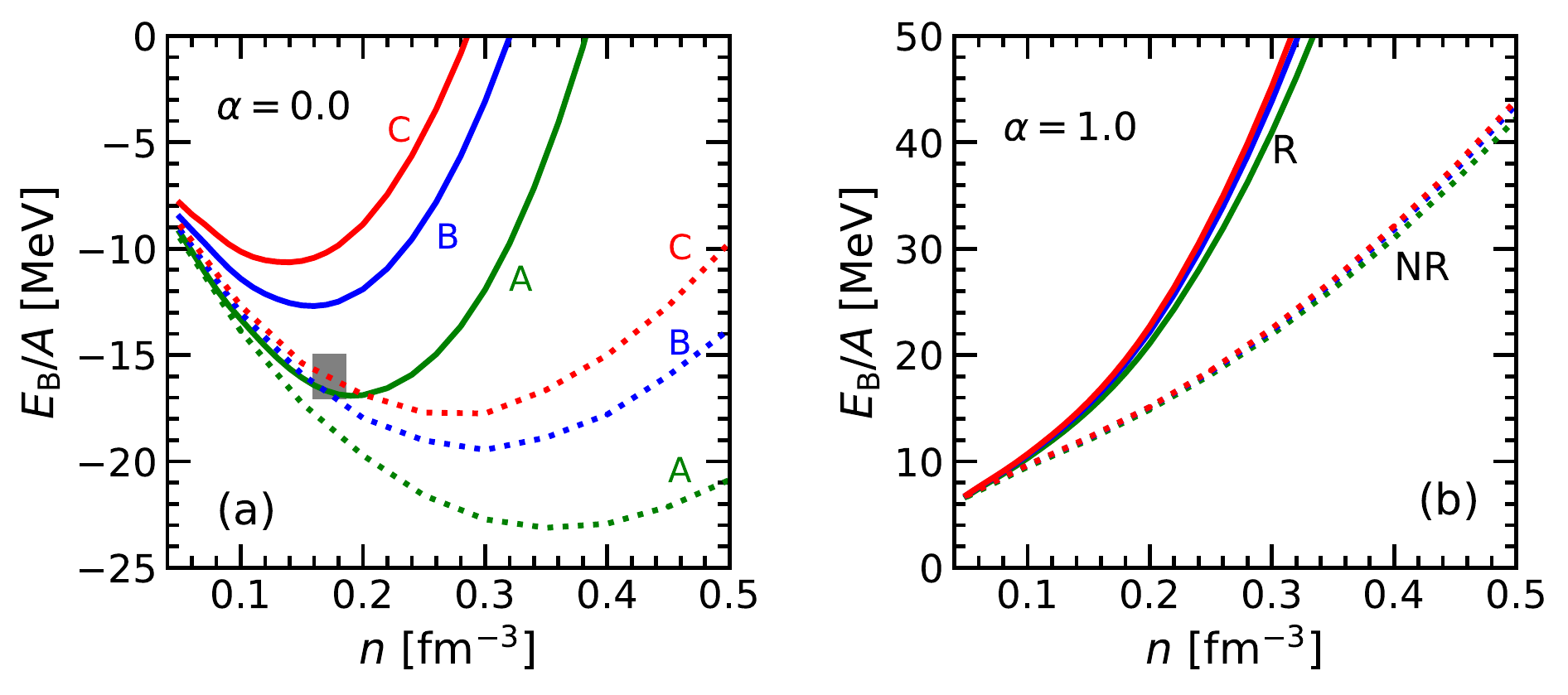}\\
	\caption{The binding energy per nucleon in symmetric nuclear matter (panel (a)) and pure neutron matter (panel (b)) with pvCDBonn potentials in the BHF and RBHF methods.}
	\label{fig5}
\end{figure}

In Fig~\ref{fig5}, the binding energy per nucleon as a function of density in the symmetric nuclear matter (panel (a)) and pure neutron matter (panel (b)) with pvCDBonn potentials in the framework of BHF and RBHF methods are shown~\cite{wang20}. In the low-density region, the equations of state (EOSs) from BHF and RBHF are quite similar to each other. As the density increases, the EOSs from relativistic theory become stiffer. The saturation densities of symmetric nuclear matter from the BHF model are larger than $0.3$ fm$^{-3}$, which are twice the empirical saturation density, while the saturation densities from the RBHF model are very close to the empirical regions. Especially, the binding energy per particle $E_B/A$ at saturation density from pvCDBonn A almost can reproduce the empirical data given by the shadow region due to the relativistic effect. Furthermore, the tensor component plays an essential role in determining the saturation properties. The pvCDBonn A has the lowest $D$-state probability, $P_D$, resulting in the highest bounding energy and saturation density. The situation for pvCDBonn C is the opposite. The strong correlation between the $P_D$ and nuclear saturation densities is called the ``Coester line"~\cite{coester70}.  

Due to the lack of $S-D$ channels in the pure neutron matter, which is isospin $T=1$ channel, the differences among the EOSs from pvCDBonn A, B, and C potentials at the same framework are very small. There are still additional repulsive contributions in the RBHF model compared to the EOSs from the BHF model. It is rapidly rising, with the density increasing. 
	
\begin{table}[htb]\centering
	\caption{The saturation properties from pvCDBonn A, B, and C potentials. 
		$n_{sat},~E_{sat},~K_\infty,~E_{sym},~L$ are nuclear saturation density, binding energy per nucleon, incompressibility, symmetry energy, and the slope of symmetry energy. Their empirical values are also given.}
	\begin{tabular}{cccccc}
		\hline\hline
		&  $n_{sat}~$[fm$^{-3}]$   & $E_{sat}/A$ [MeV]
		&  $K_\infty(n_{sat})$ [MeV] & $E_{sym}(n_{sat})$
		[MeV]&  $L(n_{sat})$ [MeV]   \\%
		\hline
		A    & 0.189 & -16.69 & 280  & 34.90 & 77.34\\
		B    & 0.155 & -13.81 & 182  & 28.99 & 49.86\\
		C    & 0.137 & -12.42 & 144  & 26.07 & 39.01\\
		Emp.  & 0.16$\pm$0.01~\cite{danielewicz09}  &-16$\pm$1~\cite{danielewicz09}
		& 240$\pm$20~\cite{garg18} & 31.7$\pm$3.2~\cite{oertel17}&
		58.7$\pm$28.1~\cite{oertel17}  \\
		\hline\hline
	\end{tabular}
	\label{tab5}
\end{table}

In Table~\ref{tab5}, the saturation properties of symmetric nuclear matter from RBHF model with pvCDBonn potentials are given. The saturation densities, $n_{sat}$ are around $0.137-0.189$ fm$^{-3}$. The saturation energies per nucleon, $E_{sat}/A$ are from $-12.42$ to $-16.69$ MeV. The incompressibilities, $K$ are about $144-280$ MeV. The symmetry energies are around $26.07-34.90$ MeV. Their slopes are $39.01-77.34$ MeV. All of them are consistent with the empirical data. 

\begin{figure}[htb]
	\centering
	\includegraphics[width=1\linewidth]{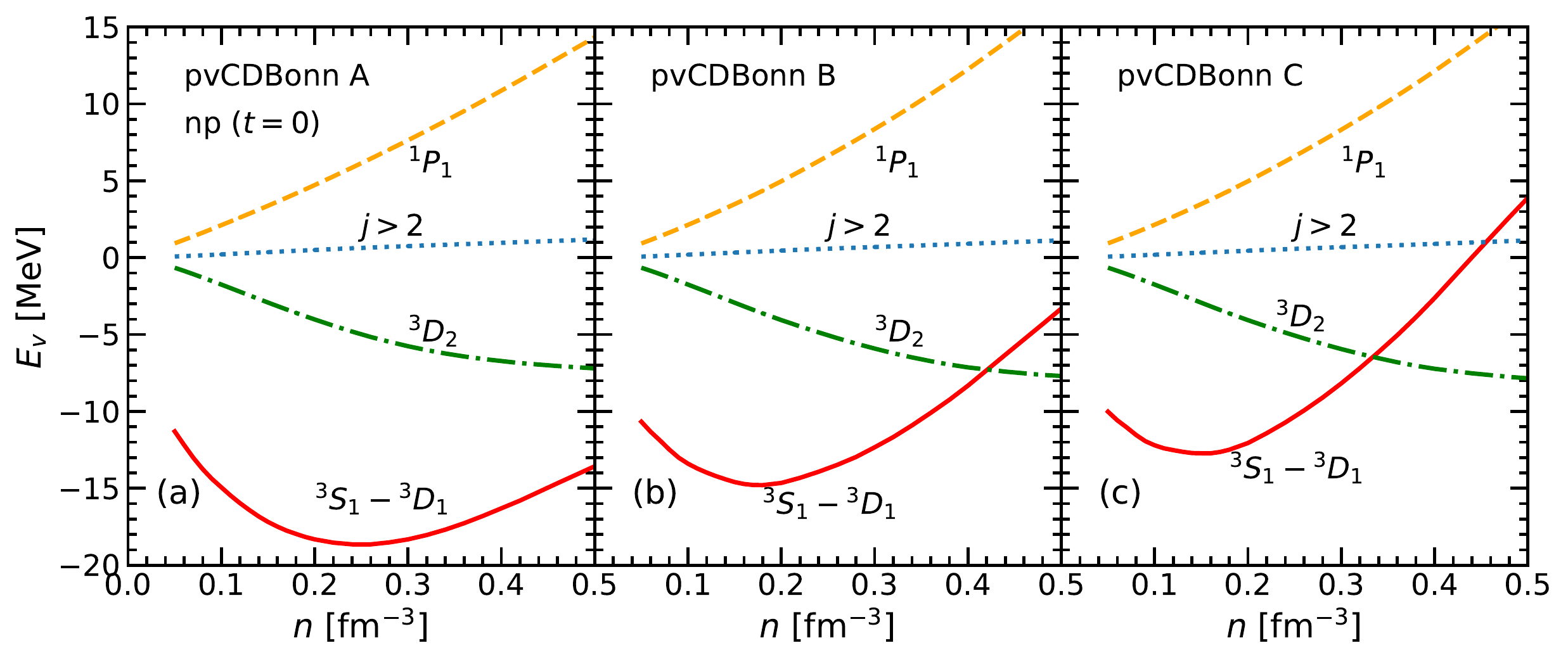}\\
	\caption{The partial wave contributions of potential energies in symmetric nuclear matter at np $T=0$ channel with pvCDBonn A, B, C potential.}
	\label{fig6}
\end{figure}

The partial wave contributions in the potential energies of symmetric nuclear matter from the $T=0$ channel as functions of density are plotted in Fig.~\ref{fig6} from pvCDBonn potentials. It is found that the contributions from $S-D$ channels have obvious differences among three potentials, which are mainly generated by the tensor force, while the energies from other channels are identical. Furthermore, there are saturation behaviors in $S-D$ channels, which are strongly correlated to the saturation density of nuclear matter.  

The symmetry energy is a very hot topic, which is determined by the isospin degree of freedom of nucleons and the Pauli principle. It can be calculated approximately with the differences in binding energy per nucleon between pure neutron matter and symmetric nuclear matter. In principle, its contribution is separated into two parts, one from kinetic energy and the other one from potential energy, with the self-energy at Fermi momentum from the Hugenholtz-Van Hove (HvH) theorem~\cite{wang22},
\begin{equation}\label{Esymdecomp}
	\begin{aligned}
		E_\mathrm{sym} & = E_\mathrm{sym}^{kin} +E_\mathrm{sym}^{pot},
		\\ E_\mathrm{sym}^{kin}& = {k_F^2 \over 4}
		\left({1+\Sigma_{{n},\mathrm{v}}\over 3E_{{n}F}^*}
		+{1+\Sigma_{{p},\mathrm{v}}\over3E_{{p}F}^*}\right)_{\alpha=0}, \\
		E_\mathrm{sym}^{pot} & = {1\over 4}\left[
		{M_{n}^*\over E_{{n}F}^*}
		{\partial \Sigma_{{n},\mathrm{s}}\over \partial \alpha}
		-{M_{p}^*\over E_{{p}F}^*}
		{\partial \Sigma_{{p},\mathrm{s}}\over \partial \alpha}+
		{k_F^2\over E_{{n}F}^*}
		{\partial \Sigma_{{n},\mathrm{v}}\over \partial \alpha}-
		{k_F^2\over E_{{p}F}^*}
		{\partial \Sigma_{{p},\mathrm{v}}\over \partial \alpha} -
		{\partial \over \partial \alpha}
		(\Sigma_{{n},0}-\Sigma_{{p},0})\right]_{\alpha=0},
	\end{aligned}
\end{equation}

\begin{figure}[htb]
	\centering
	\includegraphics[width=1\linewidth]{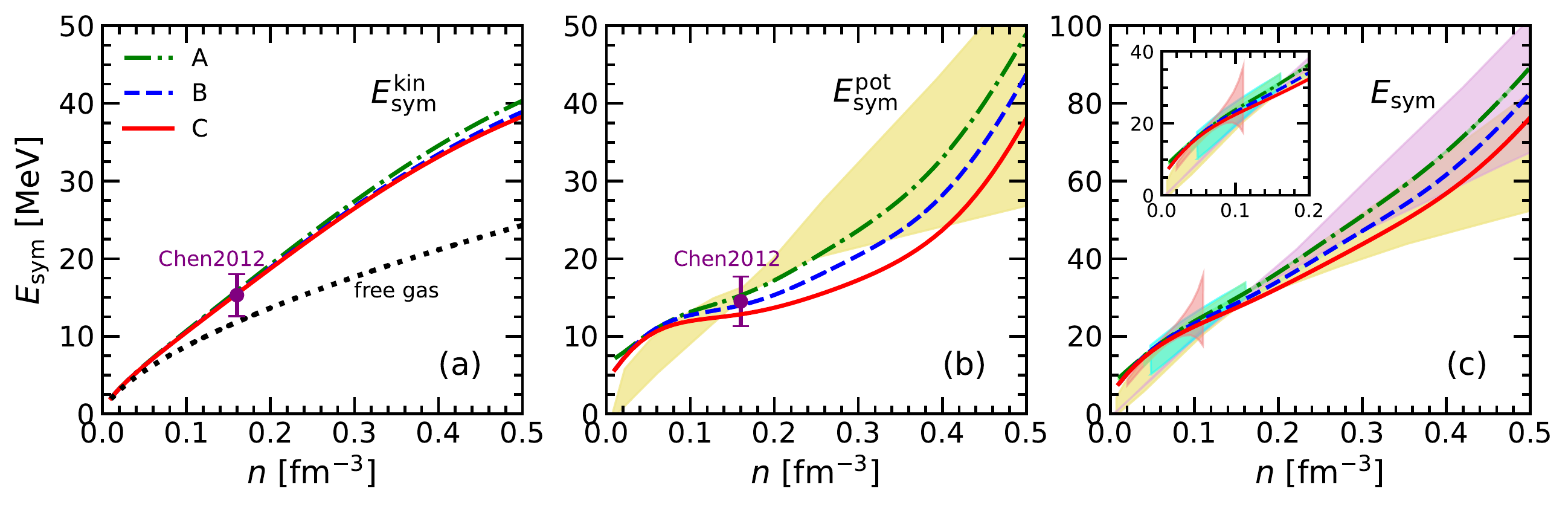}\\
	\caption{The symmetry energy and its kinetic and potential components in RBHF model with HvH theorem. The error bars in panels (a) and (b) are extracted from the data of Ref.~\cite{chen12}.}
	\label{fig7}
\end{figure}
In Fig.~\ref{fig7}, the symmetry energies and their kinetic and potential components from pvCDBonn potentials in the RBHF model are shown and compared to the analysis of the present experimental data with density functional theories. In panel (a), the kinetic energy contributions from the three potentials are almost the same and larger than the one from the free gas, since the effective nucleon mass will reduce the denominator part in Eq.~(\ref{Esymdecomp}). In panel (b), the components of symmetry energy from potential energy are exhibited. In the low-density region, their differences increase with density since the tensor components of the three potentials are different, which affects the binding energy per nucleon of symmetric nuclear matter and does not influence that of pure neutron matter. In panel (c), the total symmetry energies as functions of density are given from pvCDBonn potentials. The pvCDBonn C potential has the smallest symmetry energy and whose binding energy per nucleon of symmetric nuclear matter is the largest among the three potentials. The symmetry energies from the RBHF model are completely consistent with the experimental constraints from the heavy-ion collision, electric dipole polarizability, isobaric analog state, and the isospin diffusion experiments~\cite{danielewicz14,tsang09,tsang12,zhang15}.  

A natural extension of the investigation of nuclear matter is the study of the neutron star, which can be considered as a uniform nuclear matter consisting of neutrons, protons, electrons, and muons. They are in the conditions of $\beta$-equilibrium and charge neutrality. Therefore, the EOSs of neutron star matter with pvCDBonn potentials are calculated in the framework of the RBHF model, which uses these EOSs to evaluate the mass-radius relation of the neutron star~\cite{wang20a}. 

\begin{figure}[htb]
	\centering
	\includegraphics[width=0.5\linewidth]{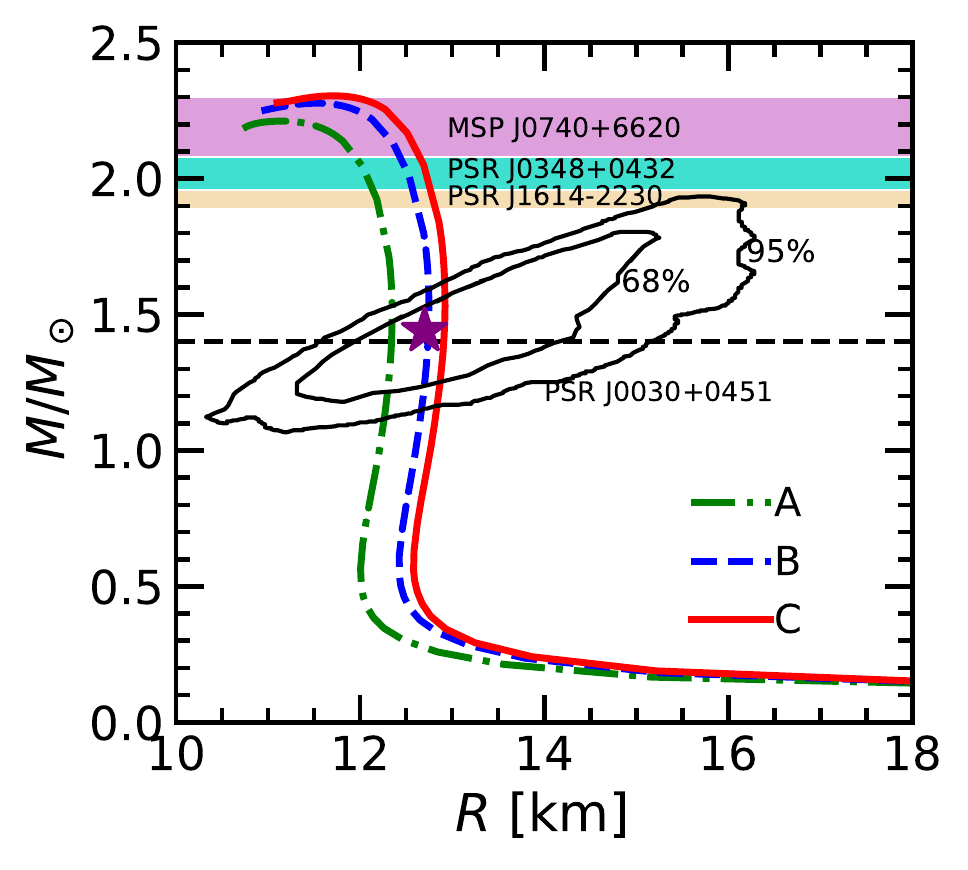}\\
	\caption{The mass-radius relation of neutron star from RBHF model with pvCDBonn potentials and compared to the constraints from the mass-radius simultaneous measurement, NICER with the confidence levels, $68\%$ and $95\%$, and the massive neutron stars, PSR J1614-2230~\cite{demorest10,fonseca16},
	PSR	J0348+0432~\cite{antoniadis13}, PSR J0740+6620~\cite{cromartie20}.}\label{fig8}
\end{figure}
In Fig.~\ref{fig8}, the mass-radius of neutron stars with pvCDBonn potentials from the RBHF model are plotted and compared with the recent constraints about the neutron star masses and radii. From 2010, three massive neutron stars were observed, PSR J1614-2230~\cite{demorest10,fonseca16}, PSR J0348+0432~\cite{antoniadis13},  PSR J0740+6620~\cite{cromartie20}, whose masses around $2M_\odot$. Further, the mass and radius of PSR J0030+0451 were simultaneously measured by the NICER collaboration group. It is found that the properties of the neutron star from the RBHF model with the high-precision NN potential satisfy these latest observables very well.
	
\section{Several new relativistic ab initio methods }
\subsection{The Hartree-Fock with UCOM model}
When the realistic NN potential is applied to the nuclear many-body system, two essential characters should be properly treated, i.e., the short-range repulsion and the tensor force in the intermediate range. Both of them are related to the high-momentum behavior of nucleons, and cannot be solved in the independent-particle models. Therefore, the beyond mean-field approximation should be developed. 
In 1998, a very powerful method with a unitary operator was proposed by Feldmeier et al., named the unitary correlation operator method (UCOM) to consider the strong repulsion at short range and the tensor component of NN interaction~\cite{feldmeier98,neff03}. The main idea of UCOM is to act as a unitary transformation on the trivial wavefunction $\phi$,
\beq
\psi=U\phi,
\eeq
to obtain an exact wave function $\psi$ for the many-body system. The corresponding operator $U$ is expressed as
\beq
U=\exp\{-iC\},~~~~C=C^\dag.
\eeq
Here $C$ is a hermitian generator including the short-range or tensor correlations. Therefore, a new equation of motion about nucleons is obtained,
\beq
&&\mathcal H\psi=E\psi,\nn
&&U^\dag\mathcal H U\phi=E\phi.
\eeq
In principle, $C$ should contain a two-body operator, three-body operator, ..., and so on, while the one-body operator only does the unitary transformation on single-particle states,
\beq
C=\sum^A_{i<j} c(i,j)+\text{three-body}+...
\eeq
An approximation of the generator should be made to remove the many-body operators except for the two-body operator~\cite{feldmeier98} since the probability of three nucleons entering the two-body short range is small at the nuclear saturation density. Therefore, a two-body correlation operator $u(i,j)$ is used to replace the generalized correlation operator $U$. For a Hamiltonian consisting of the one-body kinetic energy and the two-body potential,
\beq
\mathcal H=\sum^A_i T_i+\sum^A_{i<j}V(i,j),
\eeq
after the treatment of UCOM, it will be transferred to
\beq
\widetilde{\mathcal H}&=&u^\dag(i,j)\mathcal H u(i,j)\nn
&=&\sum^A_i T_i+\sum^A_{i<j}\widetilde V(i,j),
\eeq
where an effective NN nuclear force $\widetilde V(i,j)$ is provided with the UCOM correlations on two-body kinetic energy and the realistic NN potential,
\beq
\widetilde V(i,j)=u^\dag(i,j)V u(i,j)+u^\dag(i,j)(T_i+T_j) u(i,j)-(T_i+T_j).
\eeq
This effective interaction has a similar role to the G-matrix generated by the BHF model. However, in the real calculation, the operator form is not convenient to be used. In this work, the short-range correlation is just considered with UCOM. The $u(i,j)$ should be parameterized with a coordinate transformation function $R_+(r)$ as a function of relative distance, $r$, between two nucleons,
\beq\label{ucomfun}
R_+(r)=r+\alpha\left( \frac{r}{\beta} \right)^{\eta} \exp(-\exp(r/\beta)).
\eeq
The parameter $\alpha$ decides the overall amount of the shift and parameter, and $\beta$ is the length scale. $\eta$ represents the steepness around $r=0$. This function just acts on the short-range interaction due to the double-exponential. Therefore, the most common operators in Hamiltonian with UCOM are defined as,
\beq
u^\dag(i,j) r u(i,j)&=& R_+(r)\nn
u^\dag(i,j) V(r) u(i,j)&=& V(R_+(r))\nn
u^\dag(i,j) p_r u(i,j)&=& \frac{1}{\sqrt{R'_+(r)}}\frac{1}{r}p_r \frac{1}{\sqrt{R'_+(r)}},
\eeq
where the radial momentum is $p_r$,  and it denotes $\bra \vec r|p_r|\phi\ket=-i\frac{\partial}{\partial r}\bra \vec r|\phi\ket$. The operators are relevant to the angular momentum, as $\vec l$ and $\vec s$ are unchanged if only considering the short-range correlation.

First, the UCOM to pure neutron matte was applied in the framework of the non-relativistic Hartree-Fock model with the two and three-body NN potentials, which is called the HFUT model~\cite{hu12},
\beq
\mathcal{H}=\sum^A_{i=1}T_i+\sum^A_{i<j}V_{ij}+\sum^A_{i<j<k}V^{2\pi}_{ijk}+\sum^A_{i<j<k}V^{R}_{ijk}.
\eeq
The three nucleon potentials, $V^{2\pi}_{ijk}$ and $V^{R}_{ijk}$ correspond to the Urbana three-body interaction. Finally, the energy per particle in the HFU model with three-body interaction is obtained:
\beq\label{ehfut}
\frac{\mathcal{E}_{HFU}}{A}=\frac{3}{10}\frac{k^2_F}{M}+\frac{1}{A}\sum^A_{i<j}\bra ij|\widetilde{V}_{ij}|ij\ket_A+\frac{1}{A}\sum^A_{i<j<k}\bra ijk|\widetilde{V}_{ijk}|ijk\ket_A~,
\eeq
where $|i\ket$ is the non-relativistic plane wave function,
\beq
|i\ket=\frac{1}{\sqrt{V}}\exp(i\bm k_i\cdot\bm r)\otimes|\chi_s\ket\otimes|\chi_t\ket.
\eeq
$|\chi_s\ket$ and $|\chi_t\ket$ represent the eigenstates of spin and isospin. The parameters in UCOM correlator, $\alpha,~\beta,~\eta$ are determined by minimizing the energy per particle of the whole system with the variational principle,
\beq
\frac{\partial^3(\mathcal{E}_{HFU}/A)}{\partial\alpha\partial\beta\partial\eta}=0.
\eeq

\begin{figure}[hbt]
	\centering
	\subfigure[]{\includegraphics[width=0.49\textwidth]{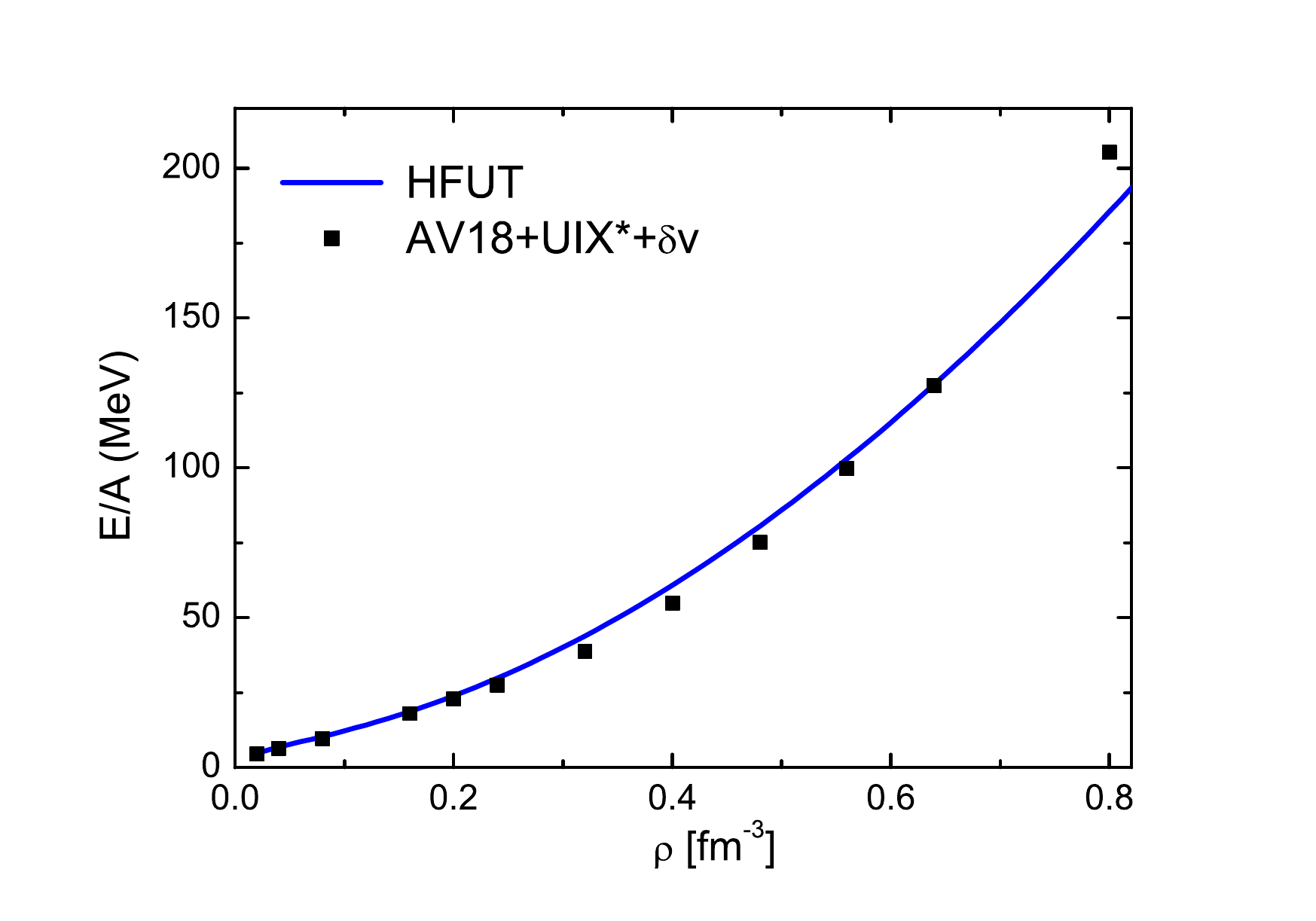}}
	\subfigure[]{\includegraphics[width=0.49\textwidth]{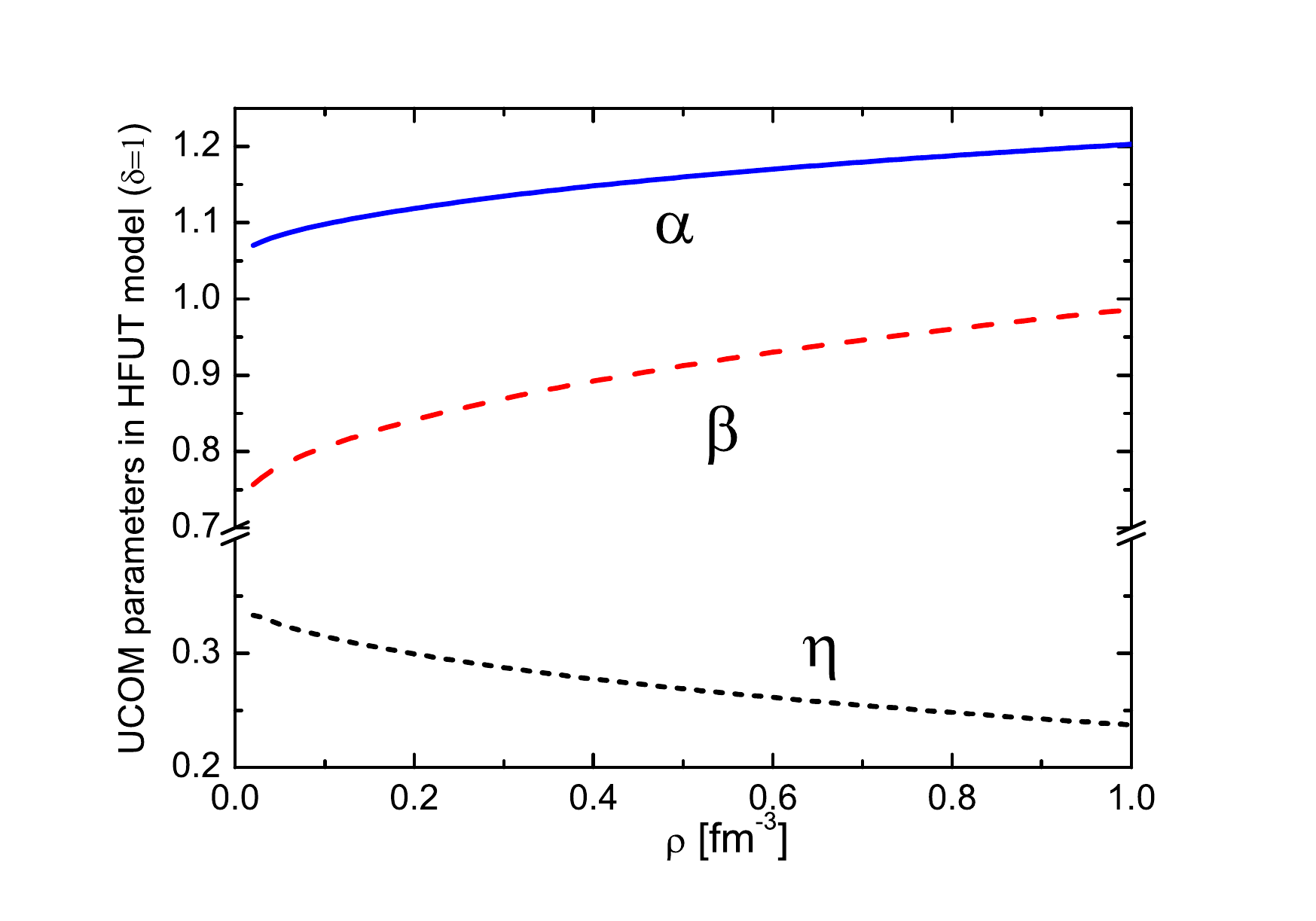}}
	\caption[The EOS of neutron matter with the HFUT model and the UCOM parameters.]{The EOS of neutron matter with the HFUT model and the UCOM parameters. In panel (a), the solid curve is the result of the HFUT model.  The square points are the EOS in the variational method with the AV18 potential and three-body interaction. In panel (b), the UCOM parameters are shown with the HFUT model for neutron matter.}\label{neos3b}
\end{figure}

The tensor effect is very weak in pure neutron matter, where it should be a good approximation to consider only the short-range correlation. The Bonn-A potential is adopted as a realistic NN interaction.
The EOS of pure neutron matter in the HFUT model is given in the (a) panel of Fig. \ref{neos3b}. The strengths of the Urbana TNI, $A_{2\pi}$ and $A_R$, are chosen as the same values as the UIX* in Ref.~\cite{akmal98}, which have the values of $A_{2\pi}=-0.0293$ MeV and $A_R=0.63\times0.048$ MeV. The result compares very well with the calculation of the variational method with the AV18 potential, including the relativistic boost correction ($\delta v$) and UIX* three-body force~\cite{akmal98}. This success is based on the following two points. The first one is that the UCOM includes a reasonable short-range correlation, and the contribution of the tensor interaction can be neglected in pure neutron matter. The other one is that the HFUT model with the Bonn potential includes the relativistic boost effect automatically. The relativistic boost correction coming from the NN potential is generated in the framework where the total momentum $\mathbf P_{ij}=\mathbf p_i+\mathbf p_j$ is zero. 

In the (b) panel of Fig.~\ref{neos3b}, the UCOM parameters, $\alpha, \beta$, and $\eta$, as a function of density in the HFUT model for pure neutron matter are given. These parameters are obtained by minimizing the ground state energy with the variational principle. The minimization of the binding energy is obtained by the competition between the short-range correlation between the kinetic energy and the potential energy. The short-range correlation effect on the kinetic energy is repulsive, while it is attractive for the potential energy. Finally, they cancel with each other and minimize the binding energy. In the high-density region, these parameters change gradually with the density and are not stabilized. This is because the three-body interaction has a large influence on the short-range correlation. In the high-density region, the repulsive contribution of the three-body interaction, for which the UCOM plays a very important role, becomes large. Recently, the similar method has been applied to study the $^4$He and neutron matter with finite particle number description~\cite{lyu20, wan20}.

A similar framework was extended to the relativistic Hartree-Fock (RHF) model~\cite{hu10}. The UCOM for relativistic kinetic energy is more complicated than the non-relativistic model due to the spinor structure, 
\beq
&&c^\dagger (i,j) T (i,j)c(i,j)-T\nn
&=&\sum_{i<j}  (\vec \alpha_i-\vec \alpha_j)\cdot \frac{\vec r}{r}\frac{1}{\sqrt{R'_+(r)}}\frac{1}{r}q_r \frac{r}{\sqrt{R'_+(r)}}\nn
&&+(\vec \alpha_i-\vec \alpha_j)\cdot \frac{\vec r}{r}\left(\frac{1}{R'_+(r)}-\frac{r}{R_+(r)}\right)q_r+\left(\frac{r}{R_+(r)}-1\right) (\vec \alpha_i-\vec \alpha_j)\cdot\vec q.
\eeq

\begin{figure}[htb]
	\centering
	\includegraphics[width=0.7\linewidth]{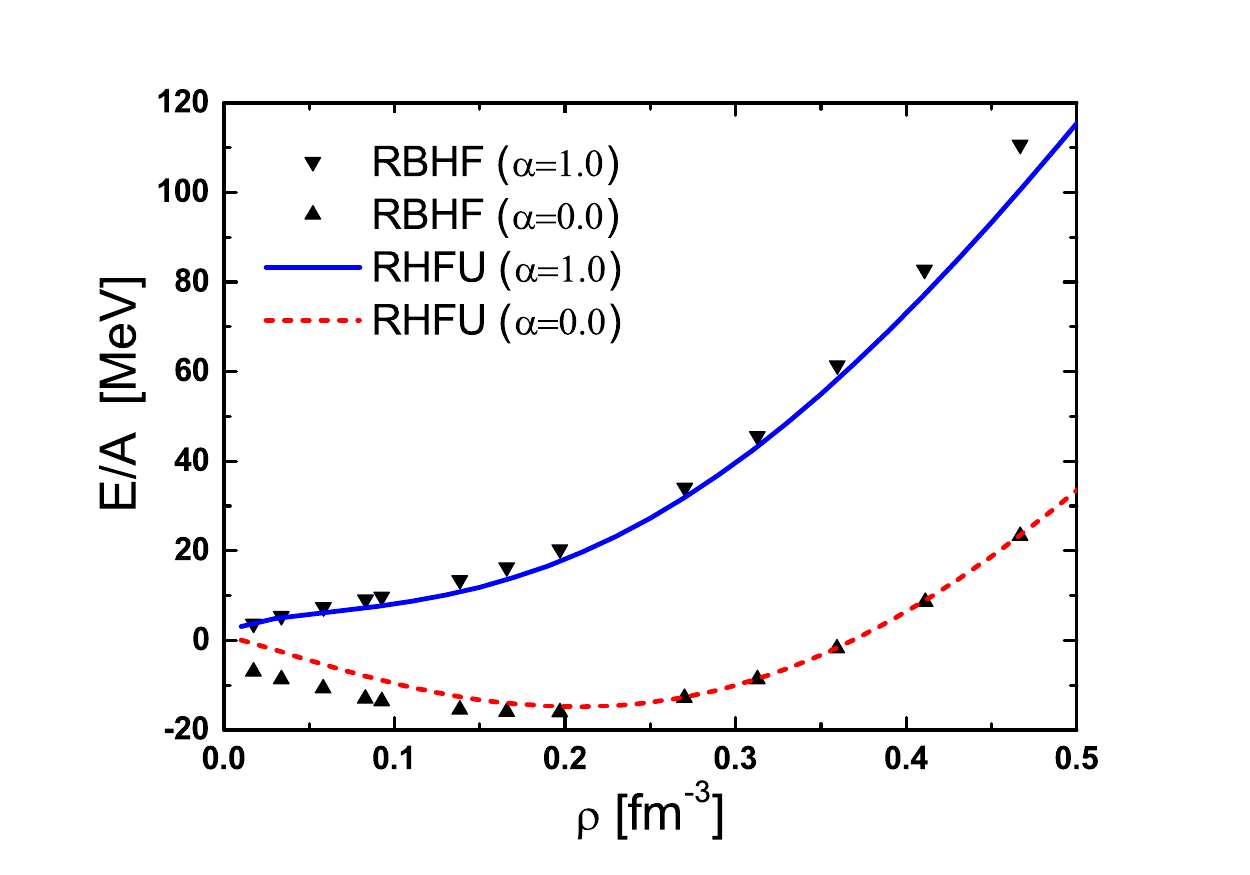}\\
	\caption{The EOSs of pure neutron matter and symmetric nuclear matter with RHFU model and Bonn A potential and comparing to the results of RBHF model.}\label{fig10}
\end{figure}
The binding energies per nucleon as functions of density for the pure neutron matter and symmetric nuclear matter with the RHF model and UCOM (RHFU) are given in Fig.~\ref{fig10}. The Bonn A potential is adopted as a realistic NN potential. The corresponding calculation results from the RBHF model are compared. The binding energies of pure neutron matter from RHFU and RBHF models are consistent with each other, similarly to the non-relativistic case. For symmetric nuclear matter, the binding energies generated by the RHFU model above $0.2$ fm$^{-3}$ accords to those from the RBHF model, while below the saturation density, the binding energy from the RHFU model is larger than that from RBHF due to the lack of tensor correlation.

\subsection{The relativistic Hartree-Fock model with high momentum components}
The tensor operator in NN potential, $S_{12}(\hat r)=\sqrt{24\pi/5}[Y_2(\hat r)\times[\sigma_1\sigma_2]^{(2)}]^{(0)}$ includes the spin operator of rank two as $[\sigma_1\sigma_2]^{(2)}$ and the spherical harmonics function $Y_2(\hat r)$. It, therefore, acts simultaneously in the spin and coordinate spaces and affects the states with different angular moments. In NN potential, the matrix elements between $S$- to $D$-states provide the largest attractive contribution to bind a proton-neutron pair and at the same time contain the high momentum components of the nucleon wave function~\cite{ikeda10}. Meanwhile, the strong shell effects of heavy nuclei are also influenced by the tensor force~\cite{toki02,myo07}.

In light nuclei, a model, called a tensor-optimized few-body model (TOFM) \cite{horii11}, assumes that the wave function of a nucleus containing only two states, $|\Psi\ket=|\Psi_S\ket+|\Psi_D\ket$ to take the tensor effect into account. This choice of considering only one $D$-state was generated by the success of the tensor optimized shell model (TOSM) applied to $^4$He by Myo et al.~\cite{myo09}.  The number of basis states in TOFM for $s$-shell light nuclei is largely reduced compared with the conventional few-body models~\cite{suzuki08, kamada01}. The results of TOFM for $A=3, 4$ systems are in good agreement with those from the stochastic variational method (SVM)~\cite{horii11}.

With this achievement in mind, a new many-body theoretical framework in the spirit of TOFM  and TOSM was proposed by Ogawa et al.~\cite{ogawa11}, where the important role of the tensor interaction is fully taken into account in the mean-field approximation. In this many-body theory, the ground state wave function is taken as a mixture of Hartree-Fock and two-particle two-hole ($2p-2h$) states, $|\Psi\ket=C_0|0\ket+\sum_\alpha C_\alpha|2p-2h,\alpha\ket$. Here, $\alpha$ distinguishes many $2p-2h$ states.  The coefficients of $2p-2h$ states and single-particle states are obtained by the variational principle of the total energy of the whole system. The equation of motion of HF single particle states was found to have a similar structure to the one in the BHF theory~\cite{ogawa11}. It is a novel theoretical framework, which can deal with the tensor force explicitly in many-body systems. Therefore, it was applied to study the properties of nuclear matter with the Bonn potentials.

In the relativistic nuclear Hamiltonian, the field operator $\psi(\mathbf x)$ can be expressed by a creation operator $c^\dag_i$ and an annihilation operator  $c_i$ of the $i$ state. Therefore, this Hamiltonian is given in a second quantization form as,
\beq
H&=&T+V\nn
&=&\sum_{\alpha\beta}\bra\alpha|T|\beta\ket c^\dag_\alpha c_\beta+\frac{1}{2}\sum_{\alpha\beta\gamma\delta}\bra\alpha\beta|V|\gamma\delta\ket c^\dag_\alpha c^\dag_\beta c_\delta c_\gamma~,
\eeq
where the indices, $\alpha,\dots$, denote the quantum numbers of momentum, spin, and isospin for all possible states. It can be separated into the particle and hole states with the particle and hole operators $a_i$ and $b_i$,
\beq
c_i=\theta(i-F) a_i +\theta(F-i) b^\dag_{\tilde i}~.
\eeq
Here, $F$ represents the states below the Fermi surface and $\theta$ function is the step function. The tilde state $\tilde i$ is a time reversal state of $i$.

The ground state wave function is assumed with the summation of HF and $2p-2h$ states,
\beq
|\Psi\ket=C_0|0\ket+\sum_\alpha C_\alpha|\alpha\ket~.
\eeq
Here, $|0\ket$ is the HF ground state, where the nucleon fully filled the states below Fermi surface and $\alpha$ denotes the quantum numbers of $2p-2h$ states,
\beq
\alpha=\{i,j,k,l\}\left\{\begin{array}{c}i,j<F~,\\ k,l>F~.\end{array}\right.
\eeq
Here, $i,j$ are hole states and $k,l$ particle states.  Therefore, the total wave function is expressed by particle and hole operators,
\beq
|\Psi\ket=C_0|0\ket+\sum_\alpha C_\alpha a^\dag_k a^\dag_l b^\dag_i b^\dag_j|0\ket~.
\eeq
For the coefficients $C_0$ and $C_\alpha$, they satisfy the normalization condition,
\beq
|C_0|^2+\sum_\alpha|C_\alpha|^2=1~.
\eeq

The total energy of the whole system can be written as with the Hamiltonian and the total wavefunction,
\beq
\bra\Psi|H|\Psi\ket&=&|C_0|^2\bra0|H|0\ket+\sum_\alpha C^*_0C_\alpha\bra0|H|\alpha\ket\nn
&&+\sum_\beta C^*_\beta C_0\bra\beta|H|0\ket+\sum_{\alpha,\beta}  C^*_\beta C_\alpha\bra\beta|H|\alpha\ket~.
\eeq
The first term is generated by the HF state.  The other three terms come from the contribution of $2p-2h$ states.  The strong tensor and short-range correlation effects are contained in these terms. Therefore, this model is called the extension of the relativistic Hartree-Fock model including the tensor correlation (RHFT).

The matrix elements in the total energy should be calculated one by one. The first one is the matrix element from the HF ground state,
\beq
\bra0|H|0\ket=\sum_{i<F}\bra i|T|i\ket+\frac{1}{2}\sum_{i,j<F}\bra ij|V|ij\ket_A~ .
\eeq
It corresponds to the standard HF energy with direct and exchange terms. The next one is matrix elements between HF and $2p-2h$ states,
\beq
\bra0|H|\alpha\ket=\bra ij|V|kl\ket-\bra ij|V|lk\ket=\bra ij|V|kl\ket_A~ ,
\eeq
where the subscript $A$ means the antisymmetrization.  Here, the kinetic energy as a one-body operator does not provide any contribution.

The final one represents the matrix elements among $2p-2h$ states.  They are very complicated and the HF energy dependent,
\beq
\bra\beta|H|\alpha\ket=\bra0|H|0\ket \delta_{\alpha,\beta}+\bra\beta|\widetilde{H}|\alpha\ket~,
\eeq
where $\bra\beta|\widetilde{H}|\alpha\ket$ are related explicitly with $2p-2h$ states.

There are two types of variational quantities in the total energy, which are the expansion coefficients $C_\alpha$ and the trivial wavefunctions $\psi_{i}(\mathbf x)$. They are determined through the variational principle of the total energy,
\beq
\frac{\delta[\bra\Psi|H|\Psi\ket-E(C^*_0C_0+\sum_\alpha C^*_\alpha C_\alpha)]}{\delta C^*_\alpha}&=&0~,\nn
\frac{\delta[\bra\Psi|H|\Psi\ket-\sum_i\varepsilon_i\int d^3\mathbf x\psi^*_i(\mathbf x)\psi_i(\mathbf x)]}{\delta\psi^*_i(\mathbf x)}&=&0~.
\eeq
The total energy $E$ and the single particle energy $\varepsilon_{i}$ are the Lagrange multipliers in the variational method.

These two variational equations can be shown explicitly as,
\beq\label{ecb}
C_0 \bra \alpha|H|0\ket + \sum_\beta C_{\beta} \bra \alpha|H| \beta \ket=E C_\alpha~,
\eeq
and
\beq\label{ebhfeq}
T|i\ket+\sum_j\bra\cdot j|V|ij\ket_A&+&C^*_0\sum_\alpha C_\alpha\frac{\partial}{\partial \psi^*_i(\mathbf x)}\bra0|H|\alpha\ket\nn
&&+\sum_{\alpha,\beta}C^*_\beta C_\alpha\frac{\partial}{\partial \psi^*_i(\mathbf x)}\bra\beta|\widetilde{H}|\alpha\ket=\varepsilon_i|i\ket~.
\eeq
In the second equation, the first two terms represent the conventional HF terms. The last two pieces are generated from contributions of $2p-2h$ states. When these two equations about coefficients and single-particle states are solved exactly, the total energy $E$ can be obtained easily.

The above modified HF equation motivates us to rewrite an effective Hamiltonian to be used for the HF state wavefunction,
\beq\label{eve}
H_{eff}=|C_0|^2 \left(H- \sum_{\alpha,\beta} V|\alpha\ket\bra \alpha|\frac{1}{H-E}| \beta\ket \bra\beta|V \right)~.
\eeq
With this effective Hamiltonian, an extended RHF differential equation for single-particle states is obtained by taking the variation of the HF matrix element of $H_{eff}$~\cite{ogawa11},
\beq\label{eve1}
\bra 0|H_{eff}|0\ket=|C_0|^2 \bra 0|T+V|0\ket - |C_{0}|^{2}\sum_{\alpha,\beta} \bra 0|V|\alpha\ket\bra \alpha|\frac{1}{H-E}| \beta\ket \bra\beta|V|0\ket~.
\eeq
The resulting differential equation is consistent with the RHFT equation (\ref{ebhfeq}), where $C_{\alpha}$ are eliminated by using the relation for the amplitudes.

The effective Hamiltonian (\ref{eve}) has a similar structure as the G-matrix of the BHF model. With some derivations~\cite{hu13,toki16}, the BHF and RHFT models can be connected, where the single-particle state in the RHFT model satisfies 
\beq
|C_{0}|^{2} \left(T+\sum_{j}\frac{\partial}{\partial\psi^*_{i}(\mathbf x) }\bra ij|\widetilde G|ij\ket_{A}\right)\psi_{i}(\mathbf x)=\varepsilon_{i}\psi_{i}(\mathbf x)~.
\eeq
The $\widetilde G$ is given by
\beq
\label{modg}
\bra ij|\widetilde G|ij\ket &=&\bra ij|V|ij\ket- \sum_{\alpha(ijp_{1}p_{2})\atop \beta(ijp'_{1}p'_{2})}\bra 0|V|\alpha \ket \bra \alpha|\frac{1}{\bra 0|H|0\ket-E+\widetilde H_{HF}+V }|\beta \ket \bra \beta|V|0\ket\nn&=&\bra ij|V|ij\ket- \sum_{\alpha(ijp_{1}p_{2})\atop\beta(ijp'_{1}p'_{2})} \bra 0|V|\alpha \ket \bra \alpha|\frac{1}{\bra 0|H|0\ket-E+\widetilde H_{HF} }|\beta \ket \bra \beta|\widetilde G|0\ket~,
\eeq
where $\widetilde H_{HF} $ is the expectation value of the Hamiltonian at Hartree-Fock approximation.

\begin{table}[!htb]
	\begin{center}
		\begin{tabular}{c c c c c c }
			\hline\hline
			{Methods}          &  {Potential} &~~ {$n_{sat}$ [fm$^{-3}$]}   &~~ {$E_{sat}/A$ [MeV]} &~~  {$K_{sat}$ [MeV]}&~~{$|C_0|^2$}\\
			\hline
			{}                 &  {Bonn A}    &~~  {0.1814}              &~~  {-15.38}     &~~  {302.9}          &~~{-}         \\
			{RBHF}             &  {Bonn B}    &~~  {0.1625}              &~~  {-13.44}     &~~  {240.3}         &~~{-}         \\
			{}                 &  {Bonn C}    &~~  {0.1484}              &~~  {-12.12}     &~~  {181.6}        &~~{-}         \\
			
			\hline
			{}                 &  {Bonn A}    &~~  {0.1699}              &~~  {-13.62}     &~~  {272.7}          &~~{0.911}     \\
			{RHFT}            &  {Bonn B}    &~~  {0.1484}              &~~  {-11.48}     &~~  {210.5}          &~~{0.905}     \\
			{}                 &  {Bonn C}    &~~  {0.1320}              &~~  { -9.80}     &~~  {163.9}          &~~{0.901}     \\
			\hline\hline
		\end{tabular}
		\caption{The saturation properties of symmetric nuclear matter in the RHFT  and RBHF models ~\cite{brockmann90} with Bonn potentials.}\label{erbhfsat}
	\end{center}
\end{table}

The saturation properties of symmetric nuclear matter, saturation density, the binding energy per nucleon, and incompressibility from the RHFT and RBHF models with Bonn A, B, and C potentials have been given in Table \ref{erbhfsat}, respectively. The binding energy per nucleon of each potential at the saturation density in the RHFT model is larger than that of the RBHF model, while the saturation densities are smaller than those of the RBHF model.  The incompressibilities of symmetric nuclear matter in the RHFT model decrease to smaller values than the RBHF results due to an additional repulsive component.  It means that the EOSs in the RHFT model are softer. Furthermore,  the binding energies at the saturation density in the RHFT model are smaller than the empirical data, around $E/A=-16\pm1$ MeV. A few MeV of attractive contribution are missing in this model and may be provided by the three-body force.

\begin{figure}[!hbt]
   \includegraphics[width=0.7\linewidth]{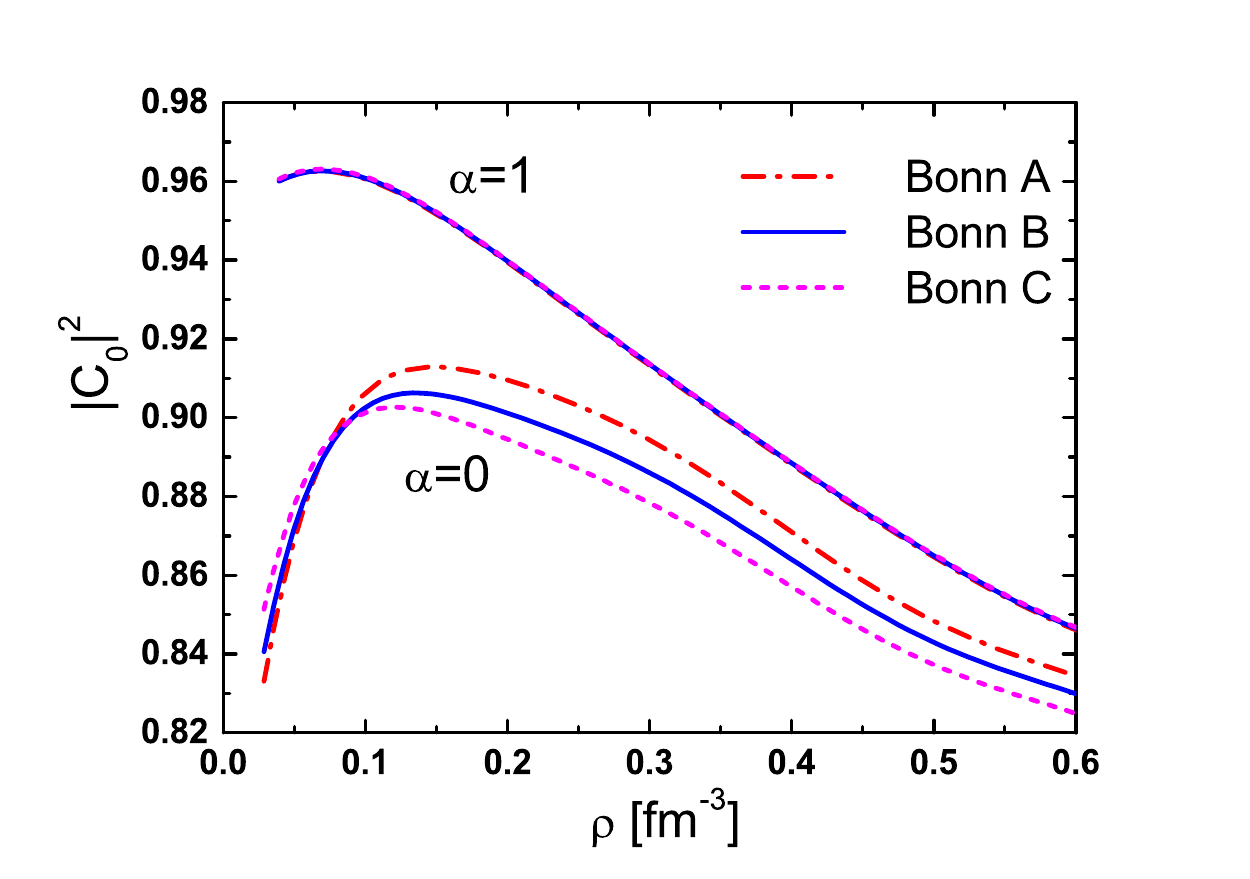}
	\caption{$|C_0|^2$ of symmetric nuclear matter and pure neutron matter in RHFT model with Bonn potentials.}\label{fig11}
\end{figure}

The components of nucleon wave function, Hartree-Fock ground states, and $2p-2h$ states are also investigated. The quantity $|C_0|^2$ can be considered as the probability of the Hartree-Fock state. The other part, $\sum_\alpha |C_\alpha|^2=1-|C_0|^2$, corresponds the components of $2p-2h$ states. $|C_0|^2$ of symmetric nuclear matter and pure neutron matter in the RHFT model with the Bonn potentials are given in Fig. \ref{fig11}. The Hartree-Fock ground states play a denominating role in the nuclear wave function since $|C_0|^2$ is above $80\%$ in the whole density region, which means that the $2p-2h$ states only provide about $20\%$ contribution to the total wave function. However, this $20\%$ takes the short-range and tensor correlations on realistic NN potential and provides the saturation mechanism of symmetric nuclear matter. Furthermore, it is also noticed that the tensor correlation plays an important role in the low-density region. The probability $|C_0|^2$ increases at low density in symmetric nuclear matter. This behavior demonstrates that the effect of the $2p-2h$ correlation becomes stronger at low density. The Bonn C potential has the largest tensor component, therefore it needs more $2p-2h$ states to include the tensor force. For the pure neutron matter, their $|C_0|^2$ are almost identical, since the tensor force, especially from the $S-D$ channels is not taken into account.

The density distribution $n(k)$ is an essential quantity to denote the high momentum components due to short-range and tensor correlations. In the HF theory, the density distribution is described by the step function,
\beq
n(k)=\left\{\begin{aligned}
	1~~~~~~~k<k_F~,\nn
	0~~~~~~~k>k_F~.
\end{aligned} \right.
\eeq

It implies that all the single particle states stay below the Fermi surface.  There is no probability in particle states with high momentum components. However, after introducing the $2p-2h$ excited states in the ground-state wave function in the RHFT model, this situation has completely changed. There is some possibility of states above the Fermi surface, which is generated by the $2p-2h$ states. The density distribution is no more the step function. The density distribution of hole states $n(i)$ is defined as,
\beq
n(i)&=&\bra\Psi|b_ib^\dag_i|\Psi\ket\nn
&=&1-\sum_{jkl}2C^*_\alpha C_\alpha~,
\eeq
and density distribution of particle states $n(k)$,
\beq
n(k)&=&\bra\Psi|a^\dag_ka_k|\Psi\ket\nn
&=&\sum_{ijl}2C^*_\alpha C_\alpha~,
\eeq
where the subscripts of coefficient $\alpha$ denote the different $i,j,k,l$ and $i,j<k_F$, $k,l>k_F$. Furthermore, the density distribution also should satisfy the identity
\beq
\int\frac{d^3\mathbf k}{(2\pi)^3}n(k)=\rho_B~.
\eeq
The normalization condition is
\beq\label{ddnt}
\int \frac{dk}{k_{F}} 3 \left(\frac{k}{k_F}\right)^{2} n(k)=1~.
\eeq

\begin{figure}[!hbt]
	\includegraphics[width=0.7\linewidth]{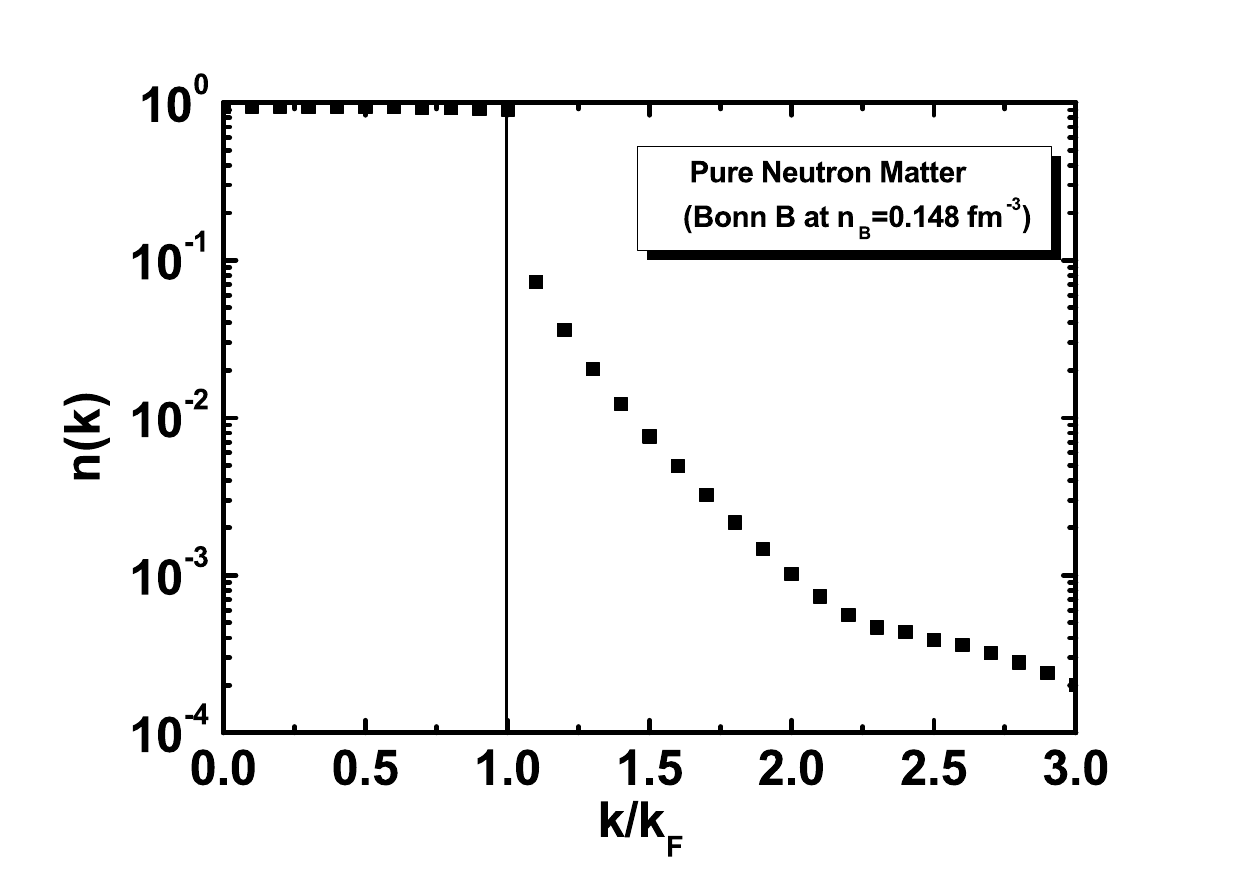}
	\caption{ The density distribution at nuclear density $n_B=0.148$ fm$^{-3}$ for symmetric nuclear matter in the RHFT model with the Bonn-B potential.}\label{fig12}
\end{figure}

Therefore, the density distribution of particle and hole states in the RHFT model with the Bonn-B potential at nuclear density $n_B=0.148$ fm$^{-3}$ is given in Fig. \ref{fig12}. It can be found that the density distribution of hole states reduces to about $0.8$ at the Fermi surface. The particle states about 0.2 close to Fermi momentum, $k_F=1.3$ fm$^{-1}$ due to the introduction of $2p-2h$ states. This result is consistent with that of the self-consistent Green's function method by Dickhoff et al. ~\cite{dickhoff04}.

\section{Summaries and perspectives} 
Due to the strong short-range repulsion and tensor force in the intermediate region, the realistic NN potential extracted from the NN scattering data is very difficult to properly describe the properties of the nuclear many-body system in the independent-particle models. Therefore, a lot of beyond-mean-field methods, i.e., {\it ab initio} methods, have been proposed recently to treat finite nuclei and infinite nuclear matter based on the non-relativistic and relativistic frameworks.

The relativistic Brueckner-Hartree-Fock (RBHF) model is a very useful {\it ab initio} method to study the nuclear many-body system, which can reproduce the saturation properties of symmetric nuclear matter very well only with the two-body realistic NN potential. However, the available realistic NN potentials suitable for the RBHF model are very few and cannot give a high-precision description of the NN scattering data. Therefore, three charge-dependent Bonn potentials with pseudovector coupling between pion and nucleon were constructed, i.e., pvCDBonn A, B, and C potentials. They include the different tensor components. They were applied to calculate the properties of nuclear matter and neutron stars. It is found that these properties are strongly correlated to the tensor components of pvCDBonn potentials.

Furthermore, several {\it ab initio} methods based on the Hartree-Fock approximation were developed to study the roles of short-range and tensor correlations of NN potential. Firstly, the unitary correlation operator method (UCOM) was adopted to study the short-range correlation in nuclear matter in the framework of the Hartree-Fock model. It is found that the properties of pure neutron matter from this model can perfectly reproduce the results from the conventional {\it ab initio} methods, while there are a few differences for the saturation properties of symmetric nuclear matter due to the lack of tensor force. Therefore, the two-particle-two-hole ($2p-2h$) states were introduced in the wavefunction of nucleon besides the Hartree-Fock (HF) states to consider the short-range and tensor correlations simultaneously. The components of HF and $2p-2h$ states were determined by the variational principle. The results with the present framework are consistent with those from the RBHF model. With the component of the $2p-2h$ state, the density distribution of nucleons in the high momentum region can also be obtained.

Although the available  {\it ab initio} method can describe the properties of nuclear matter, the accuracy is still asked to be improved. Furthermore, it is time-consuming compared to the density functional theory. It is expected that more precise relativistic  {\it ab initio} methods are applied to study the finite nuclei and infinite nuclear matter. Furthermore, the full calculations with HF and $2p-2h$ states are not completely realized in the relativistic framework due to the complicated structures of spin, which are hopeful to be solved with the quantum computing method. It can deal with the spin state very well and quickly. In the future, the role of tensor force, mainly from pions, in the nuclear many-body system must attract more attention.

\section{Acknowledgments}
We thank Prof. H. Toki, Prof. H. Shen, Dr. Y. Ogawa, Dr. Y. Zhang, and Dr. W. Wen for fruitful discussions and collaborations.  This work was supported in part by the National Natural Science Foundation of China (Grant  Nos. 11775119 and 12175109), and the Natural Science Foundation of Tianjin (Grant  No: 19JCYBJC30800).

\end{document}